 \documentclass[journal]{IEEEtran}
 
\pdfoutput=1

\usepackage{amsmath}
\usepackage{algorithm}
\usepackage{algorithmic}
\usepackage{array}
\usepackage{mdwmath}
\usepackage{eqparbox}
\usepackage{float}
\usepackage[font=footnotesize]{subfig}
\usepackage{fixltx2e}
\usepackage{stfloats}
\usepackage{amsfonts}
\usepackage{cite}
\usepackage{amsmath}
\usepackage{bbm}
\usepackage{amssymb}
\usepackage{mathrsfs}
\usepackage{amsbsy}	
\usepackage{color}
\usepackage[font=footnotesize]{subfig}
\usepackage{graphicx,epstopdf}

\usepackage{verbatim}

\floatstyle{ruled}
\newfloat{algorithm}{tbp}{loa}
\floatname{algorithm}{Algorithm}

\long\def\hide#1{}

\usepackage{comment}

\newtheorem{theorem}{\bf Theorem}[section]

\newtheorem{lemma}{\bf Lemma}[section]

\newcommand{\droped}[1]{{\color{blue} \sout{}}}

\def\ba{\begin{array}}
\def\ea{\end{array}}
\newcommand{\beq}{\begin{equation}}
\newcommand{\eeq}{\end{equation}}
\newcommand{\bq}{\begin{eqnarray}}
\newcommand{\eq}{\end{eqnarray}}
\newcommand{\bqn}{\begin{eqnarray*}}
\newcommand{\eqn}{\end{eqnarray*}}
\newcommand{\bee}{\begin{enumerate}}
\newcommand{\eee}{\end{enumerate}}
\newcommand{\bi}{\begin{itemize}}
\newcommand{\ei}{\end{itemize}}
\newcommand{\bx}{\mathbf{x}}
\newcommand{\by}{\mathbf{y}}
\newcommand{\bz}{\mathbf{z}}
\newcommand{\bc}{\mathbf{c}}
\newcommand{\bw}{\mathbf{w}}
\newcommand{\bfq}{\mathbf{q}}
\newcommand{\bfp}{\mathbf{p}}

\newboolean{showcomments}
\setboolean{showcomments}{true}
\newcommand{\qiuyu}[1]{  \ifthenelse{\boolean{showcomments}}
{ \textcolor{blue}{(Qiuyu says:  #1)}} {}  }
\newcommand{\anwar}[1]{\ifthenelse{\boolean{showcomments}}
{ \textcolor{brown}{(Anwar says: #1)} } {} }
\newcommand{\slow}[1]{\ifthenelse{\boolean{showcomments}}
{ \textcolor{red}{(Steven says:  #1)}}{}}

\begin{document}

\title{Multipath TCP: Analysis, Design and Implementation
\thanks{A preliminary version has appeared in \cite{peng2013multipath}.}
\thanks{Q. Peng and S. H. Low are with California Institute of Technology, Pasadena, CA, 91125, USA (Email:\{qpeng,slow\}@caltech.edu)}
\thanks{A. Walid is with Bell Laboratories, Murray Hill, NJ 07974, USA (Email: anwar@research.bell-labs.com)}
\thanks{J. Hwang is with Bell Labs, Alcatel-Lucent, Seoul, Korea (Email: jh.hwang@alcatel-lucent.com)}
}

\author{Qiuyu Peng, Anwar Walid, Jaehyun Hwang, Steven H. Low}

\maketitle
\begin{abstract}
Multi-path TCP (MP-TCP) has the potential to greatly improve
application performance by using multiple paths transparently.
We propose a fluid model for a large class of MP-TCP algorithms
and identify design criteria that guarantee the existence,
uniqueness, and stability of system equilibrium.   We clarify how
algorithm parameters impact TCP-friendliness, responsiveness,
and window oscillation and demonstrate an inevitable tradeoff among these properties.
We discuss the implications of these properties on the behavior
of existing algorithms and motivate our algorithm \emph{Balia} (balanced linked adaptation) which generalizes
existing algorithms and strikes a good balance among TCP-friendliness,
responsiveness, and window oscillation.
We have implemented Balia in the Linux kernel.  We use our prototype to compare the new algorithm with existing MP-TCP algorithms.

\end{abstract}



\section{Introduction}\label{sec:intro}

Traditional TCP uses a single path through the network even though multiple paths are
usually available in today's communication infrastructure; e.g.,  most
smart phones are enabled with both cellular and WiFi access, and servers in data centers
are connected to multiple routers.
Multi-path TCP (MP-TCP) has the potential to greatly improve application performance
by using multiple paths transparently.   It is being standardized by the MP-TCP Working
Group of the Internet Engineering Task Force (IETF) \cite{IETF}.
In this paper we present a fluid model of MP-TCP and study how protocol parameters
affect structural properties such as the existence, uniqueness and stability of equilibrium, the tradeoffs among TCP friendliness, responsiveness and window oscillation. These properties motivate a new algorithm that generalizes existing MP-TCP algorithms.

Various congestion control algorithms have been proposed as an extension of TCP
NewReno for MP-TCP.
A straightforward extension is to run TCP NewReno on each subpath, e.g. \cite{EWTCP,SCTP}.
This algorithm however can be highly unfriendly when it shares a path with a
single-path TCP user.
This motivates the Coupled algorithm
which is fair because it has the same underlying utility function as TCP NewReno, e.g. \cite{FPKelly_MTCP,RSrikant}.
It is found in \cite{Damon} however that the Coupled algorithm responds slowly in a dynamic network environment. A different algorithm is proposed in
 \cite{Damon} (which we refer to as the Max algorithm) which is more responsive
 than the Coupled algorithm and still reasonably friendly to single-path TCP users.
 Recently, opportunistic linked increase algorithm (OLIA) is proposed as a variant of Coupled algorithm that is as friendly as the Coupled algorithm but more responsive \cite{khalilimptcp}. See \cite{srikant2007} for more references to early work on multi-path congestion
 control.

 Our goal is to develop structural understanding of MP-TCP algorithms so that we can
 systematically tradeoff different properties such as TCP friendliness, responsiveness,
 and window oscillation that can be detrimental to applications that require a steady
 throughput.   For single-path TCP, one can associate a strictly concave utility function
 with each source so that the congestion control algorithm implicitly solves a network
 utility maximization problem \cite{FPKelly_TCP,slow_flow, srikant2007}. The convexity of this underlying
 utility maximization guarantees the existence, uniqueness, and stability of most single-path
 TCP algorithms.
 For many MP-TCP proposals considered by IETF, it will be shown that the utility maximization interpretation fails to hold in general, necessitating the need for a different approach to understanding the equilibrium properties of these algorithms. Moreover the relations among different performance metrics, such as fairness, responsiveness and window oscillation, need to be clarified.

The main contributions of this paper are three-fold.
First we present a fluid model that covers a broad class of MP-TCP algorithms and
identify the exact property that allows an algorithm to have an underlying
utility function.   This implies that some MP-TCP algorithms, e.g., the Max algorithm \cite{Damon},
has no associated utility function.
We prove conditions on protocol parameters
that guarantee the existence and uniqueness of the equilibrium,
and its asymptotical stability.
Indeed algorithms that fail to satisfy these conditions,
e.g. the Coupled algorithm, can be unstable and can have multiple equilibria
as shown in \cite{Damon}.
Second we clarify how protocol parameters impact TCP friendliness, responsiveness,
and window oscillation and demonstrate the inevitable tradeoff among these properties.
Finally, based on our understanding of the design space, we propose {\it Balia (Balanced linked adaptation)} MP-TCP algorithm
that generalizes existing algorithms and strikes a good balance among these properties.
This algorithm has been implemented in the Linux kernel and
we evaluate its performance using our Linux prototype.

We now summarize our proposed \emph{Balia} MP-TCP algorithm.
Each source $s$ has a set of routes $r$. Each route $r$ maintains a congestion window
$w_r$ and measures its round-trip time $\tau_r$. The window adaptation is as follows:
\begin{itemize}
\item For each ACK on route $r\in s$,
\end{itemize}
\bq\label{alg:increment}
w_r\leftarrow w_r+\frac{x_r}{\tau_r \left(\sum x_k\right)^2 } \left(\frac{1+\alpha_r}{2} \right) \left(\frac{4+\alpha_r}{5}\right)
\eq
\begin{itemize}
\item For each packet loss on route $r\in s$,
\end{itemize}
\bq\label{alg:decrement}
w_r\leftarrow w_r- \frac{w_r}{2}\min\left\{\alpha_r,1.5\right\}
\eq
where $x_r :=w_r/\tau_r$ and $\alpha_r:=  \frac{\max\{x_k\}}{x_r}$.

The rest of the paper is structured as follows. In Section \ref{sec:model} we develop a fluid model
for MP-TCP and use it to model existing algorithms.   In Section \ref{sec:structure}
we prove several structural properties, focusing on design criteria that determine the
existence, uniqueness, and stability of system equilibrium, TCP-friendliness,
responsiveness, window oscillation, and an inevitable tradeoff among these properties.
  In Section \ref{sec:implications} we discuss the implications of these properties
 on existing algorithms.  This motivates our new MP-TCP
  algorithm and we explain our design rationale.  In Section \ref{sec:experiment}
  we compare the performance
  of the proposed algorithm with existing algorithms using Linux implementations
of these algorithms.
  We conclude in Section \ref{sec:conc}.

 \section{Multipath TCP model}\label{sec:model}


In this section we first propose a fluid model of MP-TCP and then use it 
 to model  MP-TCP algorithms in the literature. Unless otherwise specified, a boldface letter $\bx \in \mathbb R^n$ denotes a vector with components
$x_i$.   We use $\bx_{-i} := (x_1, \dots, x_{i-1}, x_{i+1}, \dots, x_n)$ to denote the $n-1$
dimensional vector without $x_i$ and 
$\|\bx\|_{k}:=(\sum x_i^k)^{1/k}$ to denote the $L_k$-norm of $\bx$. 
Given two vectors $\bx,\mathbf{y}\in\mathbb{R}^n$, $\bx\geq \mathbf{y}$ means $x_i\geq y_i$ for all 
components $i$.    A capital letter denotes a matrix or a set, depending on the context. 
A symmetric matrix $P$ is said to be \emph{positive (negative) semidefinite} if $\bx^TP\bx \geq 0 (\leq 0)$ 
for any $\bx$, and \emph{positive (negative) definite} if $\bx^TP\bx> 0 (< 0)$ for any $\bx\neq\mathbf{0}$.
For any matrix $P$, define $[P]^+:=(P+P^T)/2$ to be its symmetric part. 
Given two arbitrary matrices $A$ and $B$ (not necessarily symmetric),
 $A\succeq B$ means $[A-B]^+$ is positive semidefinite. 
For a vector $\bx$, $diag\{\bx\}$ is a diagonal matrix with entries given by $\bx$.

\subsection{Fluid model}
 
 Consider a network that consists of a set $L=\{1,\ldots,|L|\}$ of links with finite capacities $c_l$.
 The network is shared by a set $S=\{1,\ldots,|S|\}$ of sources. Available to  
 source $s\in S$ is a fixed collection of routes (or paths) $r$.  A route $r$ consists of a set of links $l$.
 We abuse notation and use $s$ both to denote a source and the set of 
routes $r$ available to it, depending on the context. Likewise, $r$ is used both to denote
a route and the set of links $l$ in the route. Let $R := \{ r\mid r\in s, s\in S \}$ be the collection of all routes.
Let $H\in\{0,1\}^{|L|\times|R|}$ be the routing matrix: $H_{lr}=1$ if link $l$ is in route $r$ 
(denoted by `$l\in r$'), and $0$ otherwise. 

For each route $r\in R$, $\tau_r$ denotes its round trip time (RTT). 
For simplicity we assume $\tau_r$ are constants. Each source $s$ maintains a congestion window $w_r(t)$ at time $t$ for every route $r\in s$. 
Let $x_r(t):=w_r(t)/\tau_r$ represent the sending rate on route $r$. 
Each link $l$ maintains a congestion price $p_l(t)$ at time $t$. Let $q_r(t):=\sum_{l\in L}H_{lr}p_l(t)$ 
be the aggregate price  
on route $r$. In this paper  $p_l(t)$ represents the packet loss probability at link $l$ and  $q_r(t)$ 
represents the approximate packet loss probability on route $r$.

We associate three state variables $(x_r(t), w_r(t), q_r(t))$ for each route $r\in s$. 
Let $\bx_s(t):=(x_r(t),r\in s)$, $\bw_s(t):=(w_r(t),r\in s)$, $\bfq_s(t):=(q_r(t),r\in s)$.
 Then $(\bx_s(t),\bw_s(t),\bfq_s(t))$ represents the 
corresponding state variables for each source $s\in S$. For each link $l$, let 
$y_l(t) :=\sum_{r\in R} H_{lr}x_r(t)$ be its aggregate traffic rate.


Congestion control is a distributed algorithm that adapts $\bx(t)$ and $\bfp(t)$ in a closed loop. 
Motivated by the AIMD algorithm of TCP Newreno, we model MP-TCP by 
\begin{align}
\dot{x}_r&=k_r(\bx_s)\left(\phi_r(\bx_{s})-q_r\right)_{x_r}^+ & r\in s\quad s\in S
\label{eqn:TCP_dynamics}\\
\dot{p}_l&= \gamma_l \left( y_l-c_l\right)_{p_l}^+             & l\in L,
\label{eqn:AQM_dynamics}
\end{align}
where $(a)_x^+=a$ for $x>0$ and $\max\{0,a\}$ for $x\leq 0$. 
We omit the time $t$ in the expression for simplicity. 
 \eqref{eqn:TCP_dynamics} models how sending rates are adapted in the congestion avoidance
phase of TCP at each end system and \eqref{eqn:AQM_dynamics} models how the congestion 
price is (often implicitly) updated at each link. 
The MP-TCP algorithm installed at source $s$ is specified by $(K_s,\Phi_s)$, where $K_s(\bx_s):=(k_r(\bx_s),r\in s)$ and $\Phi_s(\bx_s):=(\phi_r(\bx_s),r\in s)$.  Here $K_s(\bx_s)\geq 0$ is a vector of positive gains that determines the dynamic property of the algorithm. $\Phi_s(\bx_s)$ determines the equilibrium properties of the algorithm. 
%
%
The link algorithm is specified by $\gamma_l$, where $\gamma_l>0$ is a positive gain that determines 
the dynamic property. This is a simplified model for the RED algorithm that assumes the loss probability is
proportional to the backlog, and is used in, e.g., \cite{FPKelly_TCP,slow_flow}.

\subsection{Existing MP-TCP algorithms}
\label{subsec:ea}

We first show how to relate the fluid  model  \eqref{eqn:TCP_dynamics} to the 
window-based MP-TCP algorithms proposed in the literature.
On each route $r$ the source increases its window at the return of each ACK.
Let this increment be denoted by $I_r(\bw_s)$ where $\bw_s$ is the vector of window
 sizes on different routes of source $s$. 
 The source decreases the window on route $r$ when it sees a packet loss on route 
 $r$.   Let this decrement be denoted by $D_r(\bw_s)$. Then most loss based MP-TCP algorithms
 take the form of the following pseudo code:
 \begin{itemize}
\item For each ACK on route $r$, $w_r\leftarrow w_r+I_r(\bw_s)$.
\item For each loss on route $r$, $w_r\leftarrow w_r-D_r(\bw_s)$.
\end{itemize}
 
We now model the above pseudo codes by the fluid model \eqref{eqn:TCP_dynamics}. 
Let $\delta w_r$ be the net change to window on route $r$  in each round trip time. 
Then $\delta w_r$ is roughly
\bqn
\delta w_r&=&(I_r(\bw_s)(1-q_r)-D_r(\bw_s)q_r)w_r\\
&\approx& (I_r(\bw_s)-D_r(\bw_s)q_r)w_r
\eqn
since the loss probability $q_r$ is small. On the other hand
\bqn
\delta w_r\approx\dot{w}_r\tau_r=\dot{x}_r\tau_r^2
\eqn
Hence
\bqn
\dot{x}_r=\frac{x_r}{\tau_r}\, (I_r(\bw_s)-D_r(\bw_s)q_r)
\eqn
From \eqref{eqn:TCP_dynamics} we have 
\bq\label{eqn:convert1}\left\{
\begin{array}{rcl}
k_r(\bx_s)&=& \frac{x_r}{\tau_r} D_r(\bw_s)
\\
\phi_r(\bx_s)&=& \frac{I_r(\bw_s)}{D_r(\bw_s)}
\end{array}\right.
\eq

We now apply this to the algorithms in the literature.   We
first summarize these algorithms in the form of a pseudo-code and then use 
 \eqref{eqn:convert1} to derive parameters $k_r(\bx_s)$ and $\phi_r(\bx_s)$ of
  the fluid model  \eqref{eqn:TCP_dynamics}.

 \subsubsection*{Single-path TCP (TCP-NewReno)}
 
 Single-path TCP is a special case of MP-TCP algorithm with $|s|=1$.
Hence $x_s$ is a scalar and we identify each source with its route $r=s$. 
TCP-NewReno adjusts the window as follows:
\begin{itemize}
\item For each ACK on route $r$, $w_r\leftarrow w_r+1/w_r$.
\item For each loss on route $r$, $w_r\leftarrow w_r/2$.
\end{itemize}
From \eqref{eqn:convert1}, this can be modeled by the fluid model  \eqref{eqn:TCP_dynamics}
with
\bqn
k_r (x_s) =\frac{1}{2}x_r^2, \ \ \phi_r(x_s) = \frac{2}{\tau_r^2 x_r^2}
\eqn

We now summarize some  existing MP-TCP algorithms, all of which degenerate to 
TCP NewReno if there is only one route per source.

\subsubsection*{EWTCP \cite{EWTCP}}
EWTCP algorithm applies TCP-NewReno like algorithm on each route independently
of other routes. It adjusts the window on multiple routes as follows:
\begin{itemize}
\item For each ACK on route $r$, $w_r\leftarrow w_r+a/w_r$.
\item For each loss on route $r$, $w_r\leftarrow w_r/2$.
\end{itemize}
From \eqref{eqn:convert1}, this can be modeled by the fluid model  \eqref{eqn:TCP_dynamics} with
\bqn
k_r (\bx_s) = \frac{1}{2}x_r^2, \ \ 
\phi_r(\bx_s) = \frac{2a}{\tau_r^2 x_r^2}
\eqn
where $a> 0$ is a constant.

\subsubsection*{Coupled MPTCP \cite{FPKelly_MTCP, RSrikant}}
The Coupled MPTCP algorithm adjusts the window on multiple routes in a coordinated fashion as follows:
\begin{itemize}
\item For each ACK on route $r$, $w_r\leftarrow w_r+\frac{w_r/\tau_r^2}{(\sum_{k\in s} w_k/\tau_k)^2}$.
\item For each loss on route $r$, $w_r\leftarrow w_r/2$.
\end{itemize}
From \eqref{eqn:convert1}, this can be modeled by the fluid model  \eqref{eqn:TCP_dynamics} with
\bqn
k_r (\bx_s) = \frac{1}{2}x_r^2, \ \
\phi_r(\bx_s) = \frac{2}{\tau_r^2(\sum_{k\in s}x_k)^2}
\eqn

\subsubsection*{Semicoupled MPTCP \cite{Damon}}
The Semi-coupled MPTCP algorithm adjusts the window on multiple routes as follows:
\begin{itemize}
\item For each ACK on route $r$, $w_r\leftarrow w_r+\frac{1}{\tau_r(\sum_{k\in s} w_k/\tau_k)}$.
\item For each loss on route $r$, $w_r\leftarrow w_r/2$.
\end{itemize}
From \eqref{eqn:convert1}, this can be modeled by the fluid model  \eqref{eqn:TCP_dynamics} with
\bqn
k_r (\bx_s) = \frac{1}{2}x_r^2, \ \
\phi_r(\bx_s)=  \frac{2}{x_r\tau_r(\sum_{k\in s}x_k)}
\eqn

\subsubsection*{Max MPTCP \cite{Damon}}
The Max MPTCP algorithm adjusts the window on multiple routes as follows:
\begin{itemize}
\item For each ACK on route $r$, $w_r\leftarrow w_r+
	\min \left\{ \frac{\max\{w_k/\tau_k^2\}}{(\sum w_k/\tau_k)^2},\frac{1}{w_r} \right\}$.
\item For each loss on route $r$, $w_r\leftarrow w_r/2$.
\end{itemize}
From \eqref{eqn:convert1}, this can be modeled by the fluid model  \eqref{eqn:TCP_dynamics} with
\bqn
k_r (\bx_s) = \frac{1}{2}x_r^2, \ \
\phi_r(\bx_s)=  \frac{2\max\{x_k/\tau_k\}}{x_r\tau_r(\sum_{k\in s}x_k)^2}
\eqn
where we have ignored taking the minimum with the $1/w_r$ term since the performance is mainly 
captured by $\frac{\max\{w_k/\tau_k^2\}}{(\sum w_k/\tau_k)^2}$. 

Recently, OLIA MP-TCP algorithm \cite{khalilimptcp} is shown to achieve good performance in many scenarios. OLIA uses complicated feedback congestion control signals and cannot be modeled by \eqref{eqn:TCP_dynamics}-\eqref{eqn:AQM_dynamics}. We do, however, include OLIA in our Linux-based performance evaluation in Section \ref{sec:experiment}.

\section{Structural properties}\label{sec:structure}

Throughout this paper we assume, for all $\bx_s$, $r\in s$, $s\in S$,
 $k_r(\bx_s) > 0$ and
$\phi_r(\bx_s)=0$ only if $x_k=\infty$ for some $k\in s$.
A point $(\bx, \bfp)$ is called an {\em equilibrium} of (\ref{eqn:TCP_dynamics})--(\ref{eqn:AQM_dynamics})
if it satisfies, for all $r\in s, s\in S$ and $l\in L$,
\begin{align*}
&k_r(\bx_s) \left(\phi_r(\bx_s)-q_r\right)_{x_r}^+ =0   \\
&\gamma_l \left(y_l-c_l\right)_{p_l}^+ =0
\end{align*}
or equivalently,
\bq
x_r \geq 0, \
\phi_r(\bx_s) \leq q_r & \text{and} & \phi_r(\bx_s) = q_r \text{ if } x_r > 0
\label{eqn:existence_x}
\\
p_l \geq 0, \
y_l \leq c_l & \text{and} &  y_l = c_l \text{ if } p_l > 0
\label{eqn:existence_p}
\eq

We make two remarks.  First an equilibrium $(\bx, \bfp)$ does not depend on $K_s$, but only on
$\Phi_s$.   The design $(K_s, s\in S)$ however affects dynamic properties such as stability and
responsiveness as we show below.
Second, since $k_r(\bx_s) > 0$ and $\phi_r(\bx_s)=0$ only if $x_k=\infty$ for some $k\in s$ by assumption,
any finite equilibrium $(\bx, \bfp)$ must have $q_r>0$ for all $r$.   In the following we always
restrict ourselves to finite equilibria.

In this section we denote an MP-TCP algorithm by $(K, \Phi) := (K_s, \Phi_s, s\in S)$.
We characterize MP-TCP designs $(K, \Phi)$ that guarantee
the existence, uniqueness, and stability of system equilibrium.   We  identify design
criteria that determine TCP-friendliness, responsiveness and window oscillation and prove an inevitable tradeoff among these properties.
We discuss in the next section the implications of these structural
properties on existing algorithms. All proofs are relegated to the Appendices.

\subsection{Summary}

\begin{table}
\caption{MP-TCP algorithms}
\begin{center}
\begin{tabular}{|c|c|c|c|c|c|}
\hline
  & C0& C1&C2, C3 &C4&C5\\
\hline \hline
EWTCP& Yes& Yes & Yes & Yes&Yes \\
\hline
Coupled & Yes & Yes & No& Yes&Yes\\
\hline
Semicoupled & No & Yes & Yes & Yes&Yes\\
\hline
Max & No & Yes & Yes & Yes&Yes\\
\hline
Generalized  & No & Yes & Yes & Yes&Yes\\
\hline \hline
Theorem & \ref{thm:uf} & \multicolumn{2}{|c|}{\ref{thm:existence}, \ref{thm:stability}, \ref{thm:SR} }
					& \ref{thm:friendliness} & \ref{thm:tradeoff}  \\
\hline
\end{tabular}
\end{center}
\label{tab:algsum}
\end{table}

We first present some properties of an MP-TCP algorithm $(K, \Phi)$
that we have identified.  We then interpret them and summarize their implications.
\begin{itemize}

\item [C0:] For each $s\in S$ and each $\bx_s$, the Jacobians of $\Phi_s(\bx_s)$ is continuous and symmetric, i.e.,
\bqn
\label{eqn:utility}
\frac{\partial \Phi_s}{\partial \bx_s} (\bx_s) & = &
\left[ \frac{\partial \Phi_s}{\partial \bx_s} (\bx_s)  \right]^T
\eqn

\item [C1:] For each $s\in S$ there exists a nonnegative solution
	$\bx_s := \bx_s(\bfp)$ to \eqref{eqn:existence_x} for any finite $\bfp\geq 0$ such that
	$q_r>0$ for all $r$.  Moreover,
	\bqn
		\frac{\partial y^s_l(\bfp)}{\partial p_l}\leq0, \quad \quad \lim_{p_l\rightarrow \infty} y^s_l (\bfp)=0
	\eqn
	where $y^s_l(\bfp) := \sum_{r\in s} H_{lr} x_r(\bfp)$ is the aggregate
	traffic at link $l$ from source $s$.

\item [C2:] For each $s\in S$ and each $\bx_s$, $\Phi_s(\bx_s)$ is continuously differentiable;
	moreover the symmetric part $[\partial\Phi_s(\bx_s)/\partial \bx_s]^+$ of the Jacobian is negative definite.

\item [C3:] For each $r\in R$, $\phi_r(\bx_s) = \infty$ if and only if $x_r = 0$.
	The routing matrix $H$ has full row rank.

\item [C4:] For each $r\in s$, $s\in S$, $\sum_{j\in s}[D_s]_{jr} (\bx_s) \leq0$ where
	$D_s (\bx_s) := \left[ \frac{\partial \Phi_s}{\partial \bx_s} (\bx_s) \right]^{-1}$.

\item [C5:] For each $r\in R$ and each $\bx_{-r}$, $\lim_{x_r\rightarrow \infty}\phi_r(\bx_s)=0$.

\end{itemize}

These design criteria are intuitive and usually (but not always) satisfied; see Table \ref{tab:algsum}.

Condition C0 guarantees the existence of utility functions $U_s(\bx_s)$ that an equilibrium
$(\bx, \bfp)$ of a multipath TCP/AQM (\ref{eqn:TCP_dynamics})--(\ref{eqn:AQM_dynamics})
implicitly maximizes (Theorem \ref{thm:uf}).   It is always satisfied when there is only a single path ($|s|=1$ for all $s$) but not when $|s|>1$.

Conditions C1--C3 guarantee the existence, uniqueness, and global asymptotic stability
of the equilibrium $(\bx, \bfp)$ (Theorems \ref{thm:existence} and \ref{thm:stability}).
C1 says that the  aggregate traffic rate through a link $l$ from source $s$
decreases when the congestion price $p_l$ on that link increases, and it decreases to
$0$ as $p_l$ increases without bounds.
C2 implies that at steady state, if $\bx_s, \bfq_s$ are perturbed by $\delta \bx_s, \delta \bfq_s$
respectively, then $(\delta \bx_s)^{T} \delta \bfq_s < 0$.
In the case of single-path TCP ($|s|=1$ for all $s$), C2 is equivalent to
the curvature of the utility function $U_s(x_s)$ being negative, i.e., $U_s(x_s)$ is strictly concave.
C3 means that the rate on route $r$ is zero if and only if it sees infinite price on that route.

Condition C4 is natural and satisfied by all the algorithms considered in this paper.
It allows us to formally compare MP-TCP algorithms in terms of their TCP-friendliness
(see formal definition below):
under C1--C4, an MP-TCP algorithm $(K, \Phi)$  is more friendly if $\phi_r(\bx_s)$ is smaller
(Theorem \ref{thm:friendliness}).
The existence of $D_s$ in C4 is ensured by C2.
To interpret C4, note that Lemma \ref{lem:punique} in Appendix \ref{app:existence} implies that
$\Phi_s(\bx_s^*)=\bfq_s^*$ at equilibrium.   The implicit function theorem then implies
$\mathbf{1}^T\frac{\partial \bx_s}{\partial q_r}=\sum_{j\in s}D_{jr}$ at equilibrium for all $r\in s$.
Hence C4 says that the aggregate throughput $\mathbf{1}^T\bx_s$ at equilibrium
over all routes $r\in s$ of an MP-TCP flow is a nonincreasing function of the price $q_r$.

Condition C5 is also satisfied by all the algorithms considered in this paper.
It means that the sending rate on a route $r$ grows unbounded when the congestion price
$q_r$ is zero.   Under C1--C3, an MP-TCP algorithm $(K, \Phi)$ is more responsive
(see formal definition below) if the Jacobian of $\Phi_s(\bx_s)$ is more negative definite (Theorem \ref{thm:SR}).
C5 then implies an inevitable tradeoff: an MP-TCP algorithm that is more responsive is necessarily less TCP-friendly (Theorem \ref{thm:tradeoff}).

We now elaborate on each of these properties.

\subsection{Utility maximization}\label{subsec:num}

For single-path TCP (SP-TCP), one can associate a utility function
$U_s(x_s)\in \mathbb{R}_+\rightarrow \mathbb{R}$ with each flow $s$ ($x_s$ is a scalar and $|s|=1$) and interpret (\ref{eqn:TCP_dynamics})--(\ref{eqn:AQM_dynamics})
as a distributed algorithm to maximize aggregate users' utility, e.g. \cite{FPKelly_TCP, slow_flow, srikant2007,Duality}.
Indeed, for SP-TCP, an $(\bx, \bfp)$ is an equilibrium if and only if $\bx$ is optimal for
\begin{equation}\label{eqn:NUM.1}
\mbox{maximize } \sum_{s\in S}U_s(x_s) \quad \mbox{s.t. } y_l\leq c_l \quad l\in L
\end{equation}
and $\bfp$ is optimal for the associated dual problem.
 Here $y_l\leq c_l$ means the aggregate
traffic $y_l$ at each link does not exceed its capacity $c_l$.
In fact this holds for a much wider class of SP-TCP algorithms than those
specified by (\ref{eqn:TCP_dynamics})--(\ref{eqn:AQM_dynamics}) \cite{Duality}.
Furthermore all the main TCP algorithms proposed in the literature have strictly
concave utility functions, implying a unique stable equilibrium.

The case of MP-TCP is much more delicate: whether an underlying utility function
exists depends on the design choice of $\Phi_s$ and not all MP-TCP algorithms have one.
Consider the multipath equivalent of \eqref{eqn:NUM.1}:
\begin{equation}\label{eqn:NUM}
\mbox{maximize } \sum_{s\in S} U_s(\bx_s) \quad \mbox{s.t. }  y_l\leq c_l \quad l\in L
\end{equation}
where $\bx_s := (x_r, r\in s)$ is the rate vector of flow $s$ and
$U_s : \mathbb{R}_+^{|s|}\rightarrow\mathbb{R}$ is a concave function.
\begin{theorem} [utility maximization]
\label{thm:uf}
There exists a twice continuously differentiable and concave $U_s(\bx_s)$ such that an equilibrium $(\bx, \bfp)$ of (\ref{eqn:TCP_dynamics})--(\ref{eqn:AQM_dynamics}) solves \eqref{eqn:NUM} and its dual problem if and only if condition C0 holds.
\end{theorem}

Condition C0 is satisfied trivially by SP-TCP when $|s|=1$.  For MP-TCP ($|s|>1$),
 the models derived in Section \ref{subsec:ea} show that only EWTCP and Coupled
algorithms  satisfy C0 and have underlying utility functions.  It therefore follows
from the theory for SP-TCP that EWTCP has a unique stable equilibrium while Coupled algorithm may have multiple equilibria since its corresponding utility function is not strictly concave.
The other MP-TCP algorithms all have asymmetric Jacobian $\frac{\partial \Phi_s}{\partial \bx_s}$ and do not satisfy C0.

\subsection{Existence, uniqueness and stability of equilibrium}\label{sec:existence}

Even though a multipath TCP algorithm $(K, \Phi)$ may not have a utility maximization interpretation, a unique equilibrium
exists if conditions C1--C3 are satisfied.
\begin{theorem} [existence and uniqueness]
\label{thm:existence}
\bee
\item Suppose C1 holds.  Then (\ref{eqn:TCP_dynamics})--(\ref{eqn:AQM_dynamics})  has
	at least one equilibrium.

\item Suppose C2 and C3 hold.   Then (\ref{eqn:TCP_dynamics})--(\ref{eqn:AQM_dynamics}) has
	at most one equilibrium
\eee
\noindent Thus (\ref{eqn:TCP_dynamics})--(\ref{eqn:AQM_dynamics}) has a unique
	equilibrium $(\bx^*,\bfp^*)$ under C1--C3.
\end{theorem}

Conditions C1-C3 not only guarantee the existence and uniqueness of the equilibrium,
they also ensure that the equilibrium is globally asymptotically stable, when the gain
 $k_r(\bx_s)$ is only a function of $x_r$ itself, i.e.,  $k_r(\bx_s)\equiv k_r(x_r)$ for all $r\in R$.
 This  is satisfied by all the existing algorithms presented in Section \ref{subsec:ea}.
 \begin{theorem} [stability]
 \label{thm:stability}
Suppose C1-C3 hold and  $k_r(\bx_s)\equiv k_r(x_r)$ for all $r\in R$.
Then the unique equilibrium $(\bx^*,\bfp^*)$ is globally asymptotically stable.
In particular, starting from any initial
point $\bx(0) \in \mathbb{R}^{|R|}_+$ and $\bfp(0) \in \mathbb{R}^{|L|}_+$, the trajectory
$(\bx(t), \bfp(t))$ generated by the MP-TCP algorithm (\ref{eqn:TCP_dynamics})--(\ref{eqn:AQM_dynamics})
converges to the equilibrium $(\bx^*,\bfp^*)$ as $t\rightarrow \infty$.
\end{theorem}

Our proposed algorithm does not satisfy $k_r(\bx_s)\equiv k_r(x_r)$ even though it seems to be stable in our experiments.  This condition is only sufficient and needed in our Lyapunov stability proof; see Appendix \ref{app:stability}.
 When $k_r(\bx_s)$ depends on $\bx_s$, one can replace $k_r(x_r)$ in the
 definition of the Lyapunov function $V$ in \eqref{eq:lyapunov} with $k_r(\bx_s^*)$
 evaluated at the equilibrium and the same argument there proves that $(\bx^*,\bfp^*)$
 is (locally) asymptotically stable.    Also see Theorem \ref{thm:SR} below for
 an alternative proof of local stability.




\subsection{TCP friendliness}\label{subsec:fairness}

\begin{figure}
\centering
\includegraphics[width=0.9\linewidth]{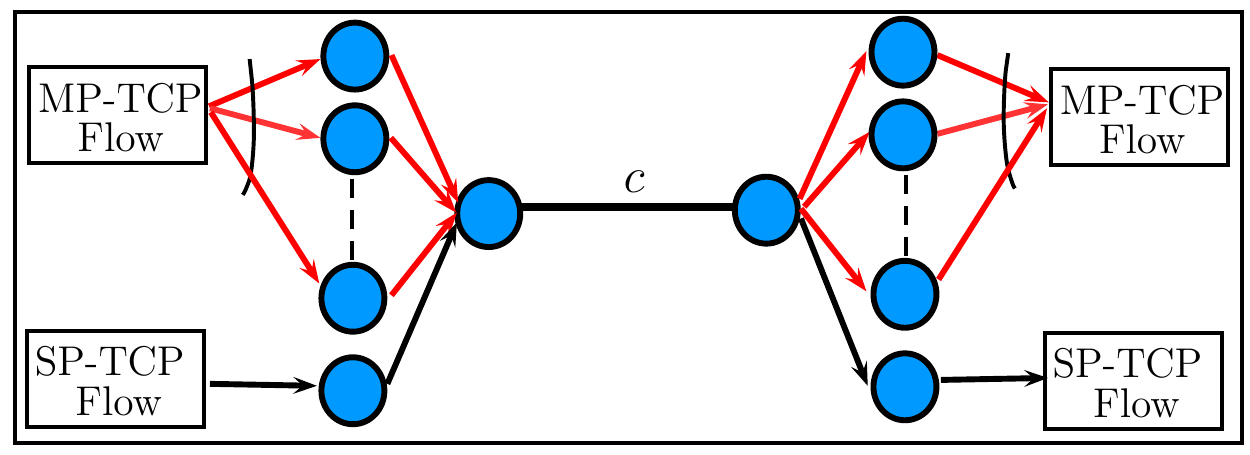}
\caption{Test network for the definition of TCP friendliness. The link in the middle is the only bottleneck link with capacity $c$. }
\label{fig:fairness_illustration}
\end{figure}

Informally, an MP-TCP flow is said to be `TCP friendly' if it does not dominate the available
bandwidth when it shares the same network with a SP-TCP flow \cite{IETF}.
To define this precisely we use the test network shared by a SP-TCP flow and a MP-TCP
flow under test as shown in Fig. \ref{fig:fairness_illustration}.
%
%

All paths traverse a single bottleneck link with capacity $c$, with all other links with capacities strictly
higher  than $c$.  The links have fixed but possibly different delays.
To compare the friendliness of two MP-TCP algorithms $\hat M := (\hat K, \hat\Phi)$
and $\tilde M := (\tilde K, \tilde\Phi)$, suppose
that when $\hat M$ shares the test network with a SP-TCP it achieves a throughput of
$\|\hat\bx\|_1$ in equilibrium aggregated over the available paths
(the SP-TCP therefore attains a throughput of $c - \|\hat\bx\|_1$).
Suppose $\tilde M$ achieves
a throughput of $\|\tilde\bx\|_1$ in equilibrium when it shares the test network with the
same SP-TCP.   Then we say that $\hat M$ is {\em  friendlier (or more TCP-friendly)}
than $\tilde M$ if $\|\hat\bx\|_1 \leq \|\tilde\bx\|_1$, i.e.,
if $\hat M$ receives no more bandwidth than $\tilde M$ does when they {\em separately} share
the test network in Fig. \ref{fig:fairness_illustration} with the same SP-TCP flow.

From the theory for single-path TCP $(|s|=1$ for all $s\in S)$, it is known that a design
is more TCP-friendly if it has a smaller marginal utility $U_s'(x_s) = \Phi_s(x_s)$.
The same intuition holds for MP-TCP algorithms
even though the utility functions may not exist for MP-TCP algorithm.
\begin{theorem} [friendliness]
\label{thm:friendliness}
Consider two MP-TCP algorithms $\hat M := (\hat K, \hat\Phi)$
and $\tilde M := (\tilde K, \tilde\Phi)$.
Suppose both satisfy C1--C4.   Then $\hat M$  is  friendlier than $\tilde M$
if $\hat\Phi_s(\bx_s)\leq \tilde{\Phi}_s(\bx_s)$ for all $s\in S$.
\end{theorem}

\subsection{Responsiveness around equilibrium}
\label{subsec:responsive}

Suppose conditions C1--C3 hold and there is a unique equilibrium $\bz^* := (\bx^*, \bfp^*)$.
Assume all  links in $L$ are active with $p_l^*>0$; otherwise remove from $L$ all
links with prices $p_l^*=0$.
Let $\delta \bz(t) := \bz(t) - \bz^*$.    The behavior of (\ref{eqn:TCP_dynamics})--(\ref{eqn:AQM_dynamics}) around the equilibrium is defined by the linearized system:
\bq
\delta \dot{\bz} & = & J^* \ \delta \bz(t)
\label{eq:ls}
\eq
Here $J^*$ is the Jacobian of (\ref{eqn:TCP_dynamics})--(\ref{eqn:AQM_dynamics})
at the equilibrium $\bz^*$:
\bqn
\label{eqn:jacobian_equilibrium}
J^*  \ := \ J(\bx^*) & := &
\begin{bmatrix}
\Lambda_{k}\frac{\partial \Phi}{\partial \bx} & -\Lambda_{k}H^T\\
  \Lambda_{\gamma}H & 0\\
\end{bmatrix}
\eqn
where $\Lambda_{k}=diag\{k_r(\bx_s^*),r\in R\}$, $\Lambda_{\gamma}=diag\{\gamma_l,l\in L\}$,
and $\frac{\partial \Phi}{\partial \bx}$ is evaluated  at $\bx^*$.

The stability and responsiveness of the linearized system \eqref{eq:ls}
(how fast does the system converges to the
equilibrium locally)  is determined by the real parts of the eigenvalues of $J^*$.
Specifically the linearized system is stable if the real parts of all  eigenvalues of
$J^*$ are negative; moreover the more negative the real parts are the faster the linearized system
converges to the equilibrium.
We now show that the linearized system \eqref{eq:ls} is stable (i.e., converges exponentially
fast to $\bz^*$ locally) and characterize its responsiveness in terms of the design
choices $(K, \Phi)$.

Let $Z=\{ \bz := (\bx, \bfp) \in \mathbb{C}^{|R| + |L|} \mid \| \bz \|_2=1 \}$.
\begin{theorem} [responsiveness]
\label{thm:SR}
Suppose C1--C3 hold.  Then
\bee
\item The linearized system \eqref{eq:ls} is stable, i.e., $\mathbf{Re} (\lambda) < 0$
	for any eigenvalue $\lambda$ of $J^*$.    Moreover
	$\mathbf{Re} (\lambda) \leq  \overline{\lambda}(J^*)$ where
	\bqn
	\overline{\lambda}(J^*) & := & \max_{ \bz \in Z}
		\left\{ \frac{\bx^H \left[ \frac{\partial \Phi}{\partial \bx} \right]^+\bx}
		{\bx^H\Lambda_{k}^{-1} \bx + \bfp^H\Lambda_{\gamma}^{-1}\bfp} \right\}
		\ \leq \ 0
	\eqn
	where $\Lambda_k$ and $\frac{\partial \Phi_s}{\partial \bx_s}$ are evaluated at
	the equilibrium point $z^*$.

\item For two MP-TCP algorithms $(\hat K, \hat \Phi)$ and $(\tilde K, \tilde \Phi)$,
	$\overline\lambda(\hat J^*)  \leq \overline\lambda (\tilde J^*)$ provided
	\bqn
	\hat K_s \ \geq \ \tilde K_s & \text{and} &
	\frac{\partial \hat\Phi_s}{\partial \bx_s}\preceq \frac{\partial \tilde\Phi_s}{\partial \bx_s}
	\qquad \text{for all } s\in S
	\eqn

\eee
\end{theorem}

Theorem \ref{thm:SR} motivates the following definition of responsiveness.
Given two MP-TCP $\hat M$ and $\tilde M$, we say that $\hat M$ is
\emph{more responsive than} $\tilde M$ if
$\overline\lambda(\hat J^*)  \leq \overline\lambda (\tilde J^*)$.
Theorem \ref{thm:SR}(2) implies that an MP-TCP algorithm with a larger $K_s(\bx^*_s)$ or more
negative definite $\left[ \frac{\partial \Phi_s}{\partial \bx_s} (\bx^*_s) \right]^+$ is more responsive, in the sense that
the real parts of the eigenvalues of the Jacobian $J^*$ have a smaller more negative upper bound.

Then the next result suggests an inevitable tradeoff between responsiveness and friendliness.
\begin{theorem} [tradeoff]
\label{thm:tradeoff}
Consider two MP-TCP algorithms  $(K, \hat\Phi)$ and $(K, \tilde{\Phi})$ with the
same gain  $K$.   Suppose both satisfy C1-C3 and C5.
Then for all $s\in S$
\bqn
\frac{\partial \hat\Phi_s(\bx_s)}{\partial \bx_s} \ \preceq \ \frac{\partial \tilde{\Phi}_s(\bx_s)}{\partial \bx_s}
& \Rightarrow &
 \hat\Phi_s(\bx_s) \ \geq \  \tilde{\Phi}_s(\bx_s)
\eqn
\end{theorem}
In light of Theorems \ref{thm:friendliness} and \ref{thm:SR}, Theorem \ref{thm:tradeoff}
says that a more responsive MP-TCP design is inevitably less friendly if they have the same $K$.

The theorem is easier to understand in the case of SP-TCP, i.e.,
when $|s|=1$ for all $s\in S$ and $\Phi_s(x_s) = U'_s(x_s)$.
 Then it implies that a more concave utility function $U_s(x_s)$ has a larger
 marginal utility, and hence less friendly.

\subsection{Window oscillation}\label{subsec:fluctuation}

Window oscillations are inherent in loss-based additive increase multiplicative decrease (AIMD) TCP algorithms.
We close this section by discussing informally why a larger design $K_s(\bx_s)$
generally creates more severe window oscillations.  This implies a tradeoff between
responsiveness (which is enhanced by a large $K_s(\bx_s))$ and oscillation
(which is reduced with a small $K_s(\bx_s))$.

The effect of $K_s(\bx_s)$ on window fluctuations can be understood by studying
how it affects the decrease $D_r(\bw_s)$ per packet loss in the following packet level model:
\begin{itemize}
\item For each ACK on route $r$, $w_r\leftarrow w_r+I_r(\bw_s)$.
\item For each loss on route $r$, $w_r\leftarrow w_r-D_r(\bw_s)$.
\end{itemize}
Let $Z_r\in\{0,1\}$ be an indicator variable of whether a packet loss is observed on
route $r$ at an arbitrary time in steady state.
Then
\bqn
D_s(\bx_s) & := & \frac{1}{\|\bx_s\|_1}\ \mathbb{E}
	\left( \, \sum_{r\in s}\frac{D_r(\bw_s)}{\tau_r}Z_r \left| \, \sum_{k\in s}Z_k\geq 1 \right. \right)
\eqn
represents the expected relative reduction in \emph{aggregate} throughput
$\sum_{r\in s}{D_r(\bw_s)}/{\tau_r}$, given that there is at least one packet loss on
some route $r\in s$.   It is a measure of throughput fluctuation for each packet loss
that an application experiences.
For TCP-NewReno (for which $s=\{r\}$ and $w_s$ is a scalar), the window size is halved on
each packet loss, $D_r(w_s)=w_r/2$, and hence $D_s(x_s)=1/2$.

To understand $D_s(\bx_s)$ for MP-TCP algorithms, we need the following result.
\begin{lemma}\label{lem:fluctuation}
Let $A_i:=\{a_{i1},a_{i2},\ldots\}$ with $|A_i|$ elements. Each element $a_{ij}$ is an
independent binary random variable with $\mathbb{P}(a_{ij}=1) = 1 - \mathbb{P}(a_{ij}=0)= q_i$.
Define $D_i(A_i):=d_i1_{(\sum_{j}a_{ij}\geq1)}$. Then
\bqn
\mathbb{E} \! \left( \! \sum_{k}D_k(A_k) \left| \, \sum_{i,j}a_{ij}\geq1 \right. \!\! \right)
	\, = \, \frac{\sum_{k}d_kq_k|A_k|}{\sum_{k}q_k|A_k|}+o\! \left( \sum_{k}q_k \right)
\eqn
\end{lemma}

Suppose each route has a fixed loss probability $q_r$. Then within each RTT, Lemma
\ref{lem:fluctuation} implies
\bqn
D_s(\bx_s) \ = \ \frac{1}{\|\bx_s\|_1}
	\left(\frac{\sum_{r\in s}w_rq_rD_r(\bw_s)/\tau_r}{\sum_{r\in s}q_rw_r}
	+ o \left(\! \sum_{r\in s}q_r \right) \! \right)
\eqn
Substituting $w_r = x_r \tau_r$ and $x_r D_r(\bw_s)= \tau_r k_r(\bx_s)$ from
\eqref{eqn:convert1}, we get,  ignoring the high-order terms,
\bq
D_s(\bx_s) & = & \frac{1}{\|\bx_s\|_1}
	\left(\frac{\sum_{r\in s}  \tau_r q_r k_r(\bx_s)}{\sum_{r\in s}\tau_r q_r x_r} \right)
\label{eqn:fluctuation_fluid}
\eq
to the first order.
Note that $k_r(\bx_s)$ does not affect the \emph{equilibrium} rates $\bx_s$.  Hence, with
the assumption that $\tau_r$ are constants, $D_s(\bx_s)$ is determined by the functions
$k_r(\bx_s)$ in steady state.

Specifically an MP-TCP algorithm with a larger $K_s(\bx_s)$ tends to have a larger
$D_s(\bx_s)$ and hence more severe window oscillations.
Theorem \ref{thm:SR} however suggests that a larger $K_s(\bx_s)$ also leads to
better responsiveness, suggesting an inevitable tradeoff between responsiveness
and window oscillation.




\section{Implications and a new algorithm}
\label{sec:implications}

In this section we discuss the implications of these structural properties
on the behavior of existing MP-TCP algorithms.   They are further illustrated
in experiment results in   Section \ref{sec:experiment}.
The discussion motivates a new design
that generalizes the existing MP-TCP algorithm.

\subsection{Implications on existing algorithms}\label{sec:511}

Recall Table \ref{tab:algsum} that summarizes the conditions satisfied by the various algorithms.
Only EWTCP and Coupled algorithms satisfy C0.  Their equilibrium properties can be studied in
the standard utility maximization model as done for single-path TCP.   Semicoupled and Max algorithms do not satisfy C0
and therefore analysis through utility maximization is not applicable.
However Theorem \ref{thm:propose}   below implies that, both Semicoupled and Max algorithms satisfy C1--C3 provided they enable no more than
$8$ routes.   Theorem \ref{thm:existence} and \ref{thm:stability} then imply that they have
a unique and globally stable equilibrium. It is also easy to show that EWTCP satisfies C1-C3.
The Coupled algorithm does not satisfy C2 and is found to have multiple equilibria in \cite{FPKelly_MTCP}.

Next we  discuss  friendliness  of existing MP-TCP algorithms.
It can be shown that the $\phi_r(\bx_s)$ corresponding to these algorithms satisfy:
\bqn
\phi_r^{ewtcp}(\bx_s)\geq\phi_r^{semicoupled}(\bx_s)\geq\phi_r^{max}(\bx_s)\geq\phi_r^{coupled}(\bx_s)
\eqn
for all $\bx_s\geq 0$ if all routes $r\in s$ have the same round trip time.
Since all of them satisfy C4, Theorem \ref{thm:friendliness} implies that their friendliness
will be in the same order, i.e., their throughputs in the test network of
Fig. \ref{fig:fairness_illustration} are ordered as follows:
\begin{center}
 EWTCP$(a\geq 1)\footnote{When $a<1$, the MP-TCP source can obtain even
 smaller throughput than the competing single-path TCP source.}
 \geq$Semicoupled$\geq$Max$\geq$Coupled
\end{center}
This is confirmed by the Linux-based experiment.

Third we will discuss responsiveness of existing MP-TCP algorithms.
These algorithms have the same gain function $k_r(\bx_s)=0.5x_r^2$ and
\bqn
(\frac{\partial \Phi_s}{\partial \bx_s})^{ewtcp}\preceq (\frac{\partial \Phi_s}{\partial \bx_s})^{semicoupled}\preceq (\frac{\partial \Phi_s}{\partial \bx_s})^{max}\preceq(\frac{\partial \Phi_s}{\partial \bx_s})^{coupled}
\eqn
Theorem \ref{thm:SR} then implies that their responsiveness should be in
the same order, as confirmed by our experiments in section \ref{sec:experiment}.

Finally we  discuss window oscillation of existing MP-TCP algorithms using
$D_s(\bx_s)$ as the metric. As mentioned in Section \ref{subsec:fluctuation}, $D_s(\bx_s)=0.5$ for
TCP NewReno, a benchmark single-path TCP algorithm. According to  \eqref{eqn:fluctuation_fluid}, if $k_r(\bx_s)\leq 0.5x_r\|\bx_s\|_1$, we have, to the first order
\bqn
D_s(\bx_s) & \leq & \frac{1}{2}\,
	\frac{\sum_{r\in s}  \tau_r q_r x_r\|\bx_s\|_1} { \|\bx_s\|_1 \sum_{r\in s}\tau_r q_r x_r}
	\ = \ \frac{1}{2}
\eqn
All existing MP-TCP algorithms have the same $k_r(\bx_s)=0.5 x_r^2\leq 0.5x_r\|\bx_s\|_1$, with strict inequality if $|s|>1$ and $x_r>0$ for at least two $r\in s$. Thus enabling MP-TCP always tends to reduce window oscillation for existing algorithms compared to TCP NewReno. Moreover, the window oscillation is always reduced compared to TCP NewReno when $k_r(\bx_s)\leq 0.5x_r\|\bx_s\|_1$.



\subsection{A generalized algorithm}

Consider the class of algorithms parametrized by $(\beta, n, \eta)$ as follows:
\bq\label{eqn:our_alg}
\left\{\begin{array}{ll}
k_r(\bx_s)&=\frac{1}{2}x_r(x_r+\eta(\|\bx_s\|_\infty-x_r)), \quad \eta\geq 0\\
\phi_r(\bx_s)&=\frac{2\left((1-\beta)x_r+\beta\|\bx_s\|_n\right)}{\tau_r^2x_r\|\bx_s\|_1^{2}}, \quad n \in \mathbb{N}_+,\beta\geq 0
\end{array}\right.
\eq
This class includes the Max $(\beta=1, \eta=0, n=\infty)$, Coupled $(\beta=0,\eta=0)$,
and Semicoupled $(\beta=1, \eta=0, n=1)$ algorithms as special cases
when all RTTs on different paths of the same source are the same,
i.e., $\tau_r = \tau_s$, $r\in s$.

The next result characterizes a subclass that have a unique and locally stable
equilibrium point.
\begin{theorem}\label{thm:propose}
Fix any $\eta\geq 0$ and $n\in \mathbb{N}_+$.
For any $s\in S$, the $\phi_r(\bx_s)$  in \eqref{eqn:our_alg} satisfies
\bee
\item   C1 if $\beta \geq 0$.
\item   C2--C3 if $0<\beta\leq1$, $|s|\leq 8$ and $\tau_r$ are the same for all $r\in s$
	(assuming $H$ has full row rank).
\eee
\end{theorem}
The requirement that $|s|\leq 8$ is not restrictive since in practice a device
may typically enable no more than 3 paths.
The requirement that $\tau_r$ are the same for all $r\in s$ is used in proving the
negative definiteness of the (symmetric part of the) Jacobian of $\Phi_s(\bx_s)$.   Since a
negative definite matrix remains negative definite after small enough perturbations
of its entries, Theorem \ref{thm:propose} holds if the RTTs of the subpaths do not differ much.
This (sufficient) condition seems reasonable
as two paths between the same source-destination pair often have similar RTTs
if both are wireline paths. Note that our experiments in Section \ref{sec:experiment} show that the algorithm also converges
even if the RTTs on different paths differ dramatically, e.g. the RTT of WiFi is usually much smaller than that of 3G.

For the class of algorithms specified by \eqref{eqn:our_alg}, Theorem \ref{thm:propose}
motivates a design space defined by $\beta\in (0, 1], \eta \geq 0, n\in \mathbb N_+$, where $\beta$ and $n$ control the tradeoff between friendliness and responsiveness and $\eta$ controls the tradeoff between responsiveness and window oscillation. In Table \ref{tab:designspace}, we summarize how the parameters $(\beta,\eta,n)$ affect the performance.

We now describe our design philosophy. As discussed above the design of MP-TCP algorithms involves inevitable tradeoffs
among responsiveness, friendliness, and the severity of window oscillation.
Specifically a design is more responsive if it has a higher gain $K_s$ or a more negative
definite Jacobian $\left[ \partial \Phi_s/\partial\bx_s \right]^+$ (Theorem \ref{thm:SR}).
However a larger $K_s$ usually creates a bigger window oscillation;
a more negative definite $\left[ \partial \Phi_s/\partial\bx_s \right]^+$ implies
a larger $\Phi_s$, usually hurting friendliness (Theorems \ref{thm:tradeoff} and
\ref{thm:friendliness}).   This is summarized in Table \ref{tab:designspace}.
Since enabling multiple paths already reduces window oscillation compared to single-path TCP (section \ref{sec:511}),
MP-TCP can afford to use a relatively large gain $K_s$ for responsiveness.
This does not compromise too much on window oscillation, but allows us to
use a less negative definite Jacobian $\left[ \partial \Phi_s/\partial\bx_s \right]^+$
with a smaller $\Phi_s$ to maintain sufficient TCP friendliness. Moreover, responsiveness is mainly affected by subpaths with small throughput
while window oscillation is mainly affected by subpaths with large throughput. The parameter $\eta$ in the generalized algorithm \eqref{eqn:our_alg} scales $k_r(\bx_s)$ in the right way: a path $r$ that has a large $x_r$ has $k_r(\bx_s) \approx 0.5 x_r^2$ and hence a similar
degree of window oscillation as existing algorithms, while  a path $r$ with a small
$x_r$ has larger $k_r(\bx_s)$ than that under a design with zero $\eta$ and therefore is
more responsive.

Our experiments show that Max algorithm ($(\beta,\eta,n)=(1,0,\infty)$) overtakes too much of the competing single-path TCP flows. Hence, we can only use a smaller $\beta$ since $n$ is already infinite in order to improve friendliness. To compensate the responsiveness performance, we will use a larger $\eta$, which will sacrifice window oscillation performance. The \emph{Balia} MP-TCP algorithm given at the end of Section \ref{sec:intro} corresponds to the choice
$(\beta,\eta,n)=(0.2, 0.5, \infty)$. Instead of allowing the window size to drop to $1$ for a packet loss, we add a cap for the decrement of window size, which improves the performance according to our experiments. Note that there is no ``best'' parameter settings since there are tradeoffs among all the performance metrics and we choose $(\beta,\eta,n)=(0.2, 0.5, \infty)$ based on our experiments in Section \ref{sec:experiment}, which show that this parameter setting strikes a good balance among responsiveness, friendliness, and window oscillation.


\begin{table}
\centering
\caption{How design choices affect MP-TCP performance.}
\begin{tabular}{|c|c|c|}
\hline
{\bf Performance} & {\bf  Parameter} & {\bf Parameters in \eqref{eqn:our_alg}} \\ 
\hline
TCP friendliness &   $\phi_r(\bx_s)\downarrow$  & $\beta\downarrow$, $n\uparrow$ \\ 
\hline
Responsiveness & $k_r(\bx_s)\uparrow$,  $-\partial\Phi_s/\partial \bx_s\uparrow$ & $\beta\uparrow$, $n\downarrow$, $\eta\uparrow$  \\ 
\hline
Window oscillation & $k_r(\bx_s)\downarrow$  & $\eta\downarrow$\\ 
\hline
\end{tabular}
\label{tab:designspace}
\end{table}

\section{Experiment}\label{sec:experiment}


\begin{figure}
\centering
\includegraphics[width=0.9\linewidth]{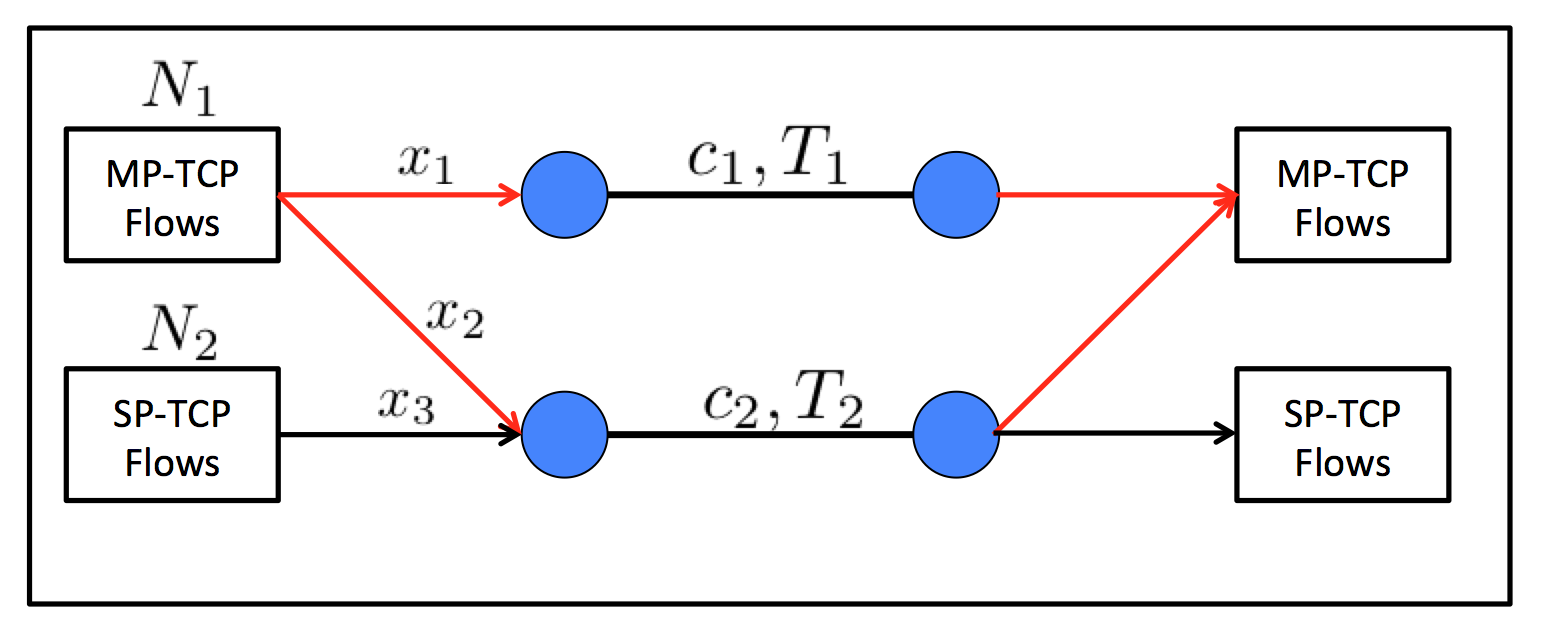}
  \caption{Network for our Linux-based experiments on TCP friendliness and responsiveness,
  with $N_1$ MP-TCP flows and $N_2$ single-path TCP flows sharing $2$ links of capacity $c_1$, $c_2$ and propagation delay (single trip) $T_1$, $T_2$. MP-TCP flows maintain two routes with rate $x_{1}$, $x_{2}$. Single-path TCP flows maintain one route with rate $x_{3}$.}
\label{fig:top1}
\end{figure}

In this section we summarize our experimental results that illustrate the above analysis. In addition to the MP-TCP algorithms illustrated in section \ref{subsec:ea}, we also include the recently developed OLIA MP-TCP algorithm \cite{khalilimptcp}. We evaluate the MP-TCP algorithms using a reference Linux implementation of MP-TCP, Multipath TCP v0.88~\cite{MPTCP}. Since it currently includes only Max and OLIA algorithms,
we implement EWTCP, Semicoupled, Coupled, and the proposed Balia algorithm in the reference implementation.
For the Coupled and our algorithm, the minimum {\it ssthresh} is set to $1$ instead of $2$
when more than $1$ path is available.

The network topology is shown in Fig.~\ref{fig:top1}. In the testbed, all nodes are 
Linux machines with a quad-core Intel i5 3.33GHz processor, 4GB RAM and multiple 1Gbps 
Ethernet interfaces, running Ubuntu 13.10 (Linux kernel 3.11.8). The network
parameters such as $c_1$, $c_2$, $T_1$, and $T_2$ are controlled by Dummynet~\cite{Dummynet}.

Our experiments are divided into three parts.
First we compare TCP friendliness of Balia algorithm and prior algorithms.
The result confirms that the Couple algorithm is the friendliest, Balia algorithm is close to the Coupled
algorithm and friendlier than the other algorithms.
Second we compare the responsiveness of each algorithm in a dynamic environment
where flows come and go.
The result shows that the Coupled and OLIA algorithms are unresponsive (illustrating the tradeoff
between responsiveness and friendliness).   EWTCP is the most responsive; Balia is
similar in responsiveness but friendlier to single-path TCP flows.
Finally we show that all MP-TCP algorithms have smaller average window oscillations than single-path TCP.

These experiments confirm our analytical results and suggest our design choice strikes a
good balance among friendliness, responsiveness, and window oscillation.

\subsection{TCP friendliness}

We study TCP friendliness of each algorithm, first with paths of similar RTTs and then with paths of different RTTs, which emulates the wireless scenario.
We assume all the flows are long lived and focus on the steady state throughput.



In the first set of experiments, we let $T_1=T_2=5$ms, $c_1=c_2=60$Mbps
and $N_1=N_2=30$. We repeat the experiments 20 times, the average aggregate throughput of MP-TCP and single-path TCP users and the $95\%$ margin of error for \emph{confidence interval} (CI) are shown in Table \ref{table:top1}.
The Coupled algorithm is the friendliest and Balia algorithm is closer to Coupled algorithm than the others. 

\begin{table}[htbp]
  \centering
    \caption{TCP friendliness (same RTTs): Average throughput (Mbps) and 95\% confidence interval of MP-TCP and single-path TCP users.
    ($T_1=T_2=5$ms, $c_1=c_2=60$Mbps and $N_1=N_2=30$)}
  \begin{tabular}{|>{\centering} m{1.7cm}|c|c|c|c|c|c|}
    \hline
           & ewtcp   & semi.  & max    & balia    & coupled & olia     \\\hline
    mp-tcp (throuput)  &2.75	 &2.65 & 2.60 &	2.52&	2.44 &2.61   \\\hline
    mp-tcp (CI)& 0.005&	0.004&0.005	&0.006	&0.005&	0.004   \\\hline
    sp-tcp (throuput) & 0.951&	1.07&1.13&	1.22&	1.29 &	1.12  \\\hline
    sp-tcp (CI)& 0.005& 0.007&	0.008 &	0.006&	0.005&	0.004 \\\hline
  \end{tabular}
\label{table:top1}
\end{table}

In the second set of experiments, we assume a highly heterogeneous RTTs by emulating the scenario of a mobile device with both 3G and WiFi access. WiFi access usually has higher capacity and lower delay compared to 3G. Specificially, we set $T_1=10$ms, $c_1=8$Mbps for the first link to emulate WiFi access and $T_2=100$ms, $c_2=2$Mbps for the second link to emulate 3G access. When there exists single-path TCP flows, i.e. $N_2>0$, the behaviors of all the algorithms are similar to the equal RTT case in the first set of simulation. The Coupled algorithm is the friendliest and Balia algorithm is closer than other algorithms. However, when there is no single-path TCP flow, i.e.  $N_1=1$ and $N_2=0$, 
the performance of OLIA is not stable to effectively take all the available capacity while the other algorithms do not have such problem. We repeat the experiments 20 times and we find sometimes  OLIA does not use the 3G access link. The average throughput of MP-TCP user and the 95\% margin of error for confidence interval  is shown in Table \ref{table:top3}.

\begin{table}[htbp]
  \centering
    \caption{Basic behavior (WiFi/3G): throughput (Mbps) of a MP-TCP user and 95\% confidence interval.
    ($T_1=10$ms, $T_2=100$ms, $c_1=8$Mbps, $c_2=2$Mbps and $N_1=1, N_2=0$)}
  \begin{tabular}{|p{1.9cm}|c|c|c|c|c|c|}
    \hline
           & ewtcp   & semi.   & max     & balia    & coupled  & olia      \\\hline
    throughput & 9.26&	9.27 &	9.26 &9.27	&9.28&	9.19   \\\hline
    {\scriptsize confidence interval } & 0.008&0.006&	0.006&	0.01 &	0.01 &	0.09   \\\hline
  \end{tabular}
\label{table:top3}
\end{table}

\subsection{Responsiveness}\label{exp:responsive}

\begin{figure}
\centering
\subfloat {
	\includegraphics[width=0.45\linewidth,height=0.35\linewidth]{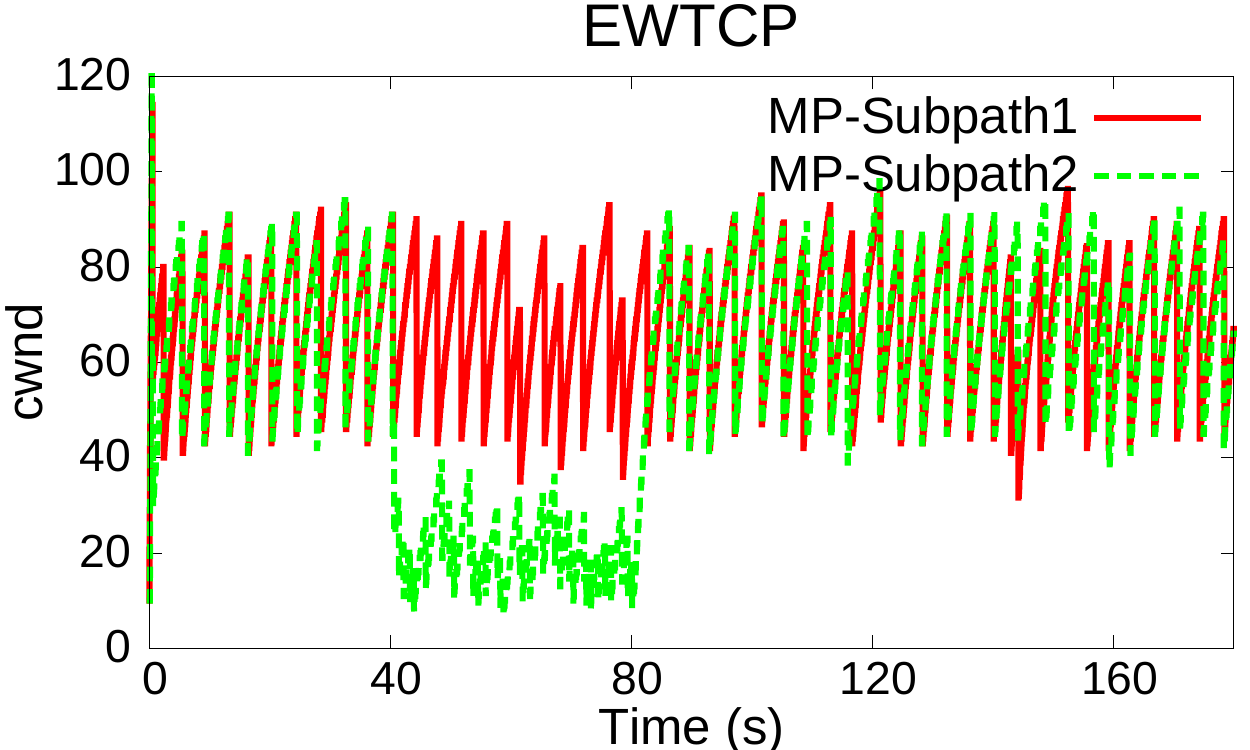}
	\label{fig:exp_responsiveness_cwnd_ewtcp}
	}
\subfloat{
	\includegraphics[width=0.45\linewidth,height=0.35\linewidth]{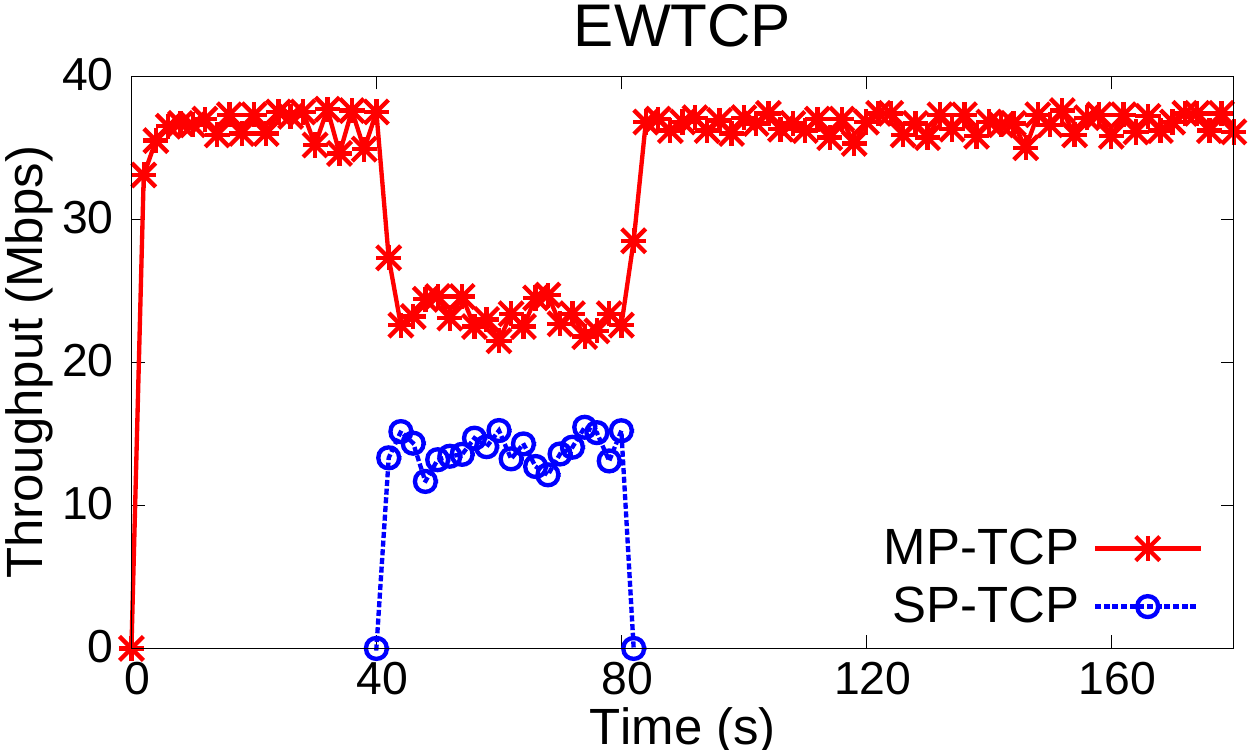}
	\label{fig:exp_responsiveness_thru_ewtcp}
	}
	\\
\subfloat{
	\includegraphics[width=0.45\linewidth,height=0.35\linewidth]{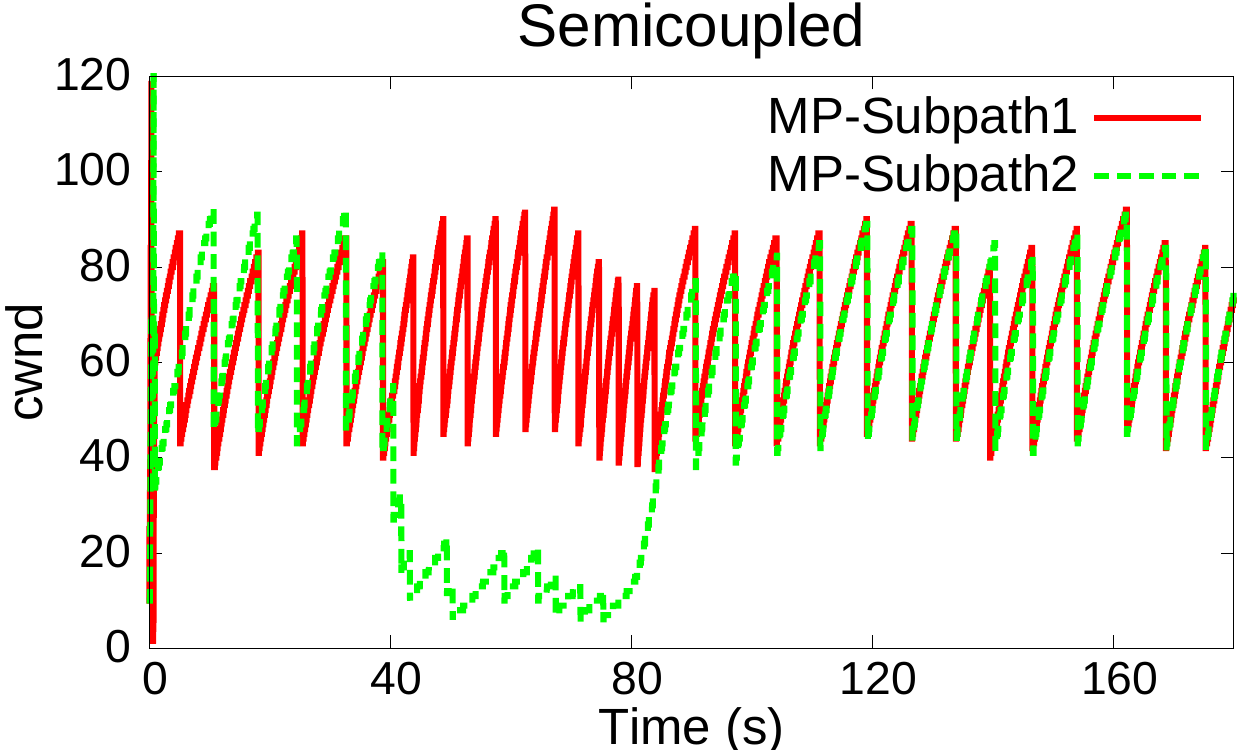}
\label{fig:exp_responsiveness_cwnd_semicoupled}
	}
\subfloat{
	\includegraphics[width=0.45\linewidth,height=0.35\linewidth]{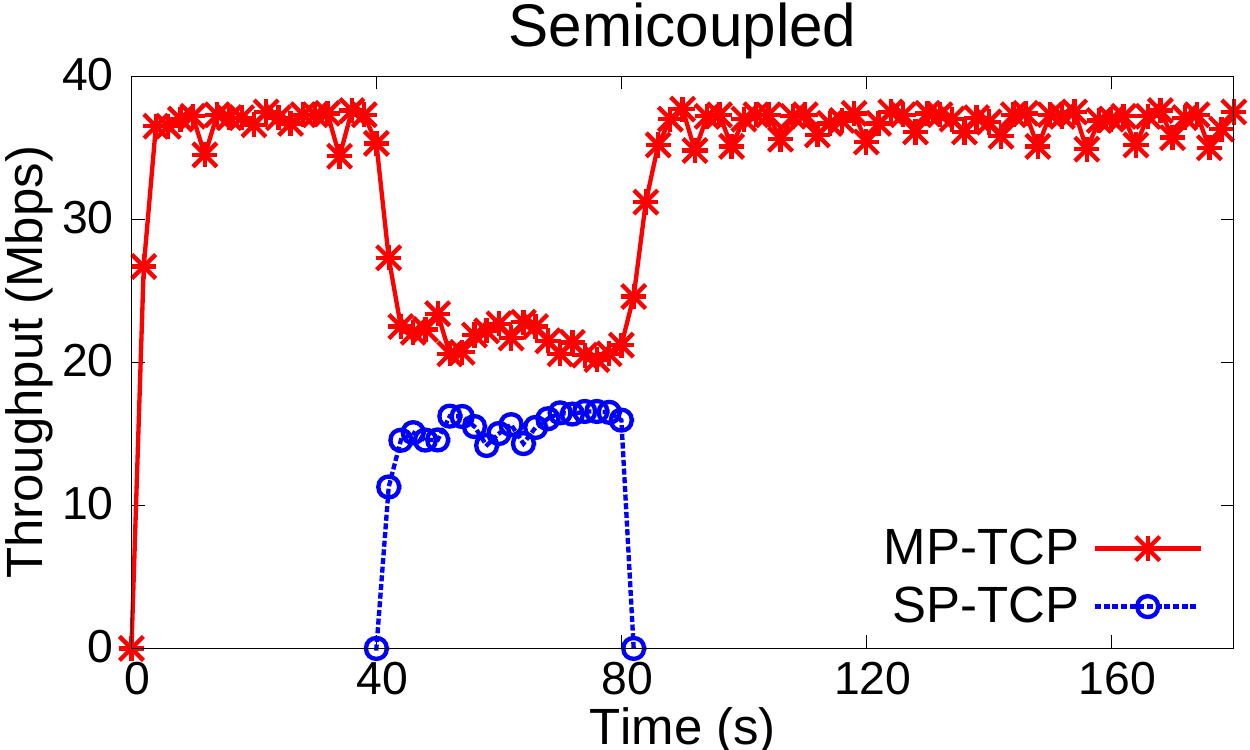}
\label{fig:exp_responsiveness_thru_semicoupled}
	}
	\\
\subfloat{
	\includegraphics[width=0.45\linewidth,height=0.35\linewidth]{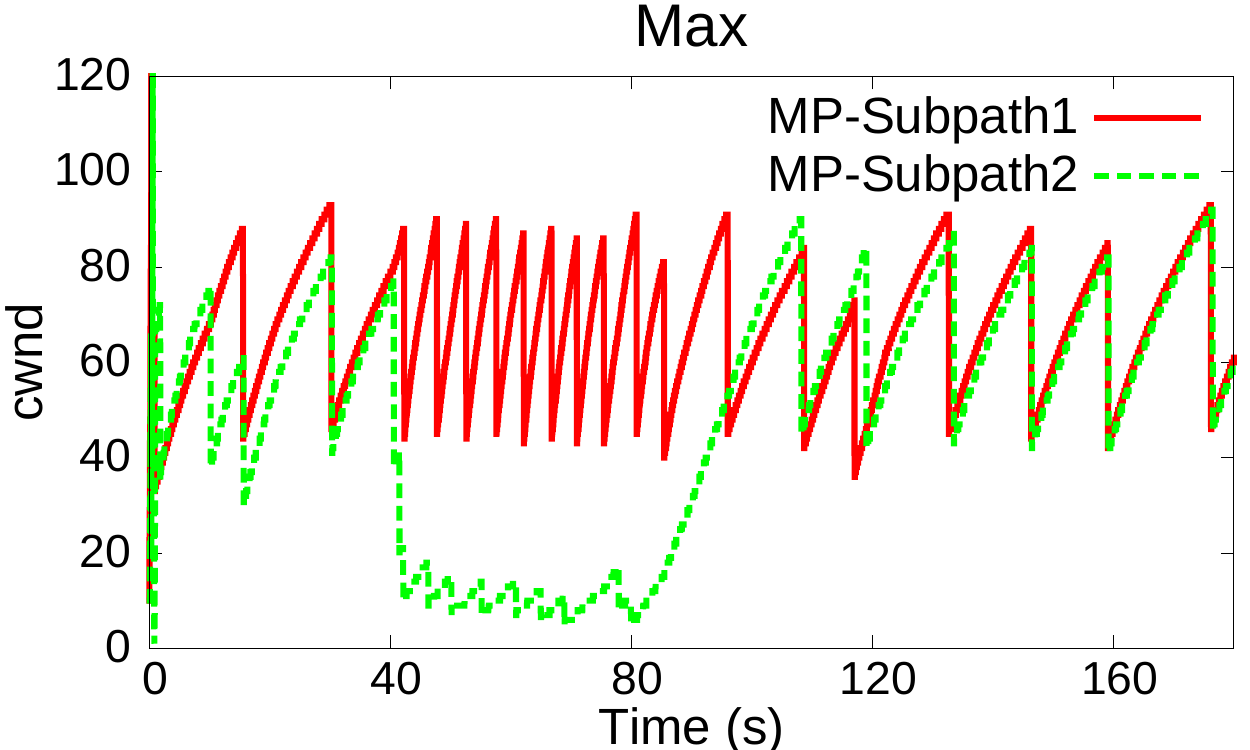}
\label{fig:exp_responsiveness_cwnd_max}
	}
\subfloat{
	\includegraphics[width=0.45\linewidth,height=0.35\linewidth]{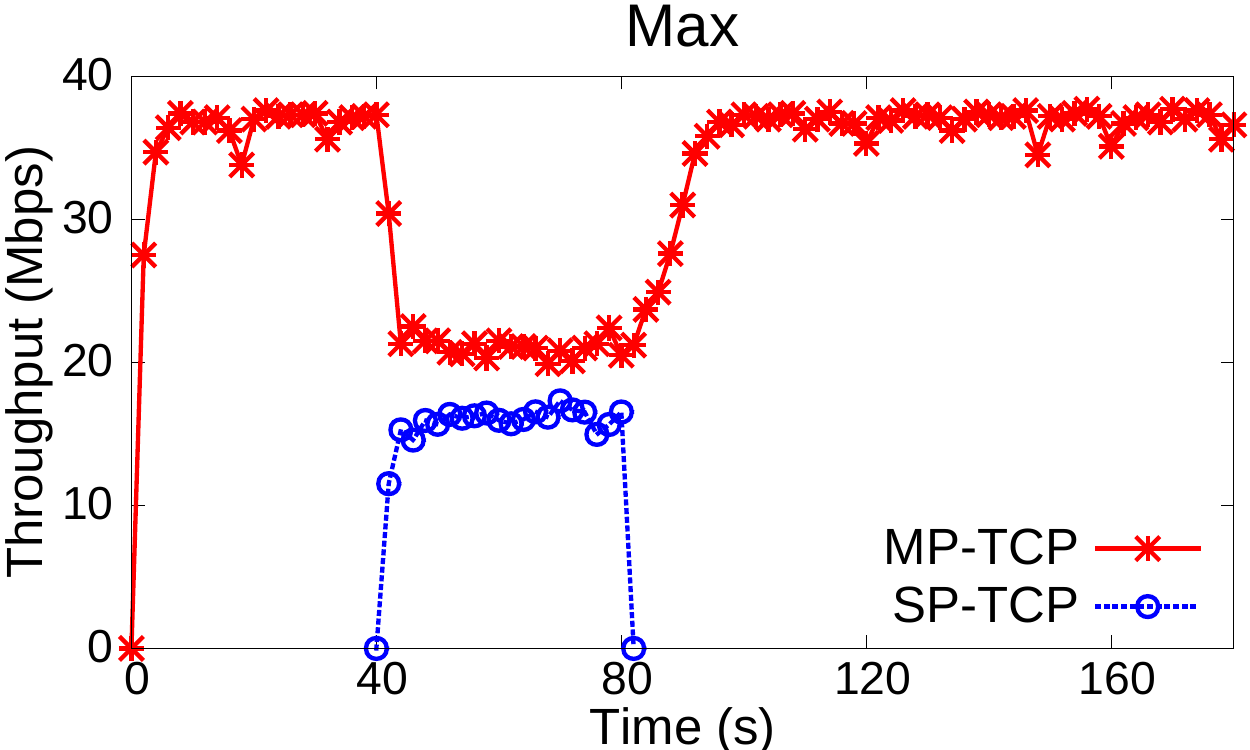}
\label{fig:exp_responsiveness_thru_max}
	}
	\\
\subfloat{
	\includegraphics[width=0.45\linewidth,height=0.35\linewidth]{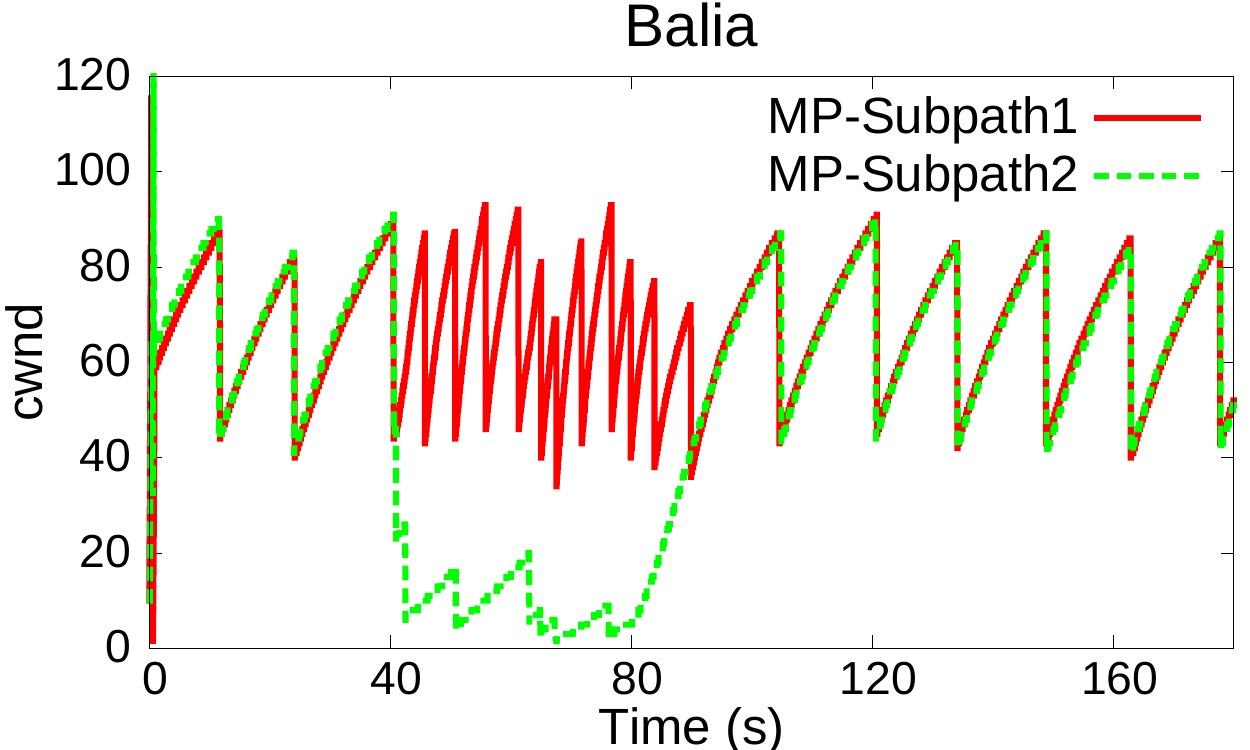}
\label{fig:exp_responsiveness_cwnd_ours}
	}
\subfloat{
	\includegraphics[width=0.45\linewidth,height=0.35\linewidth]{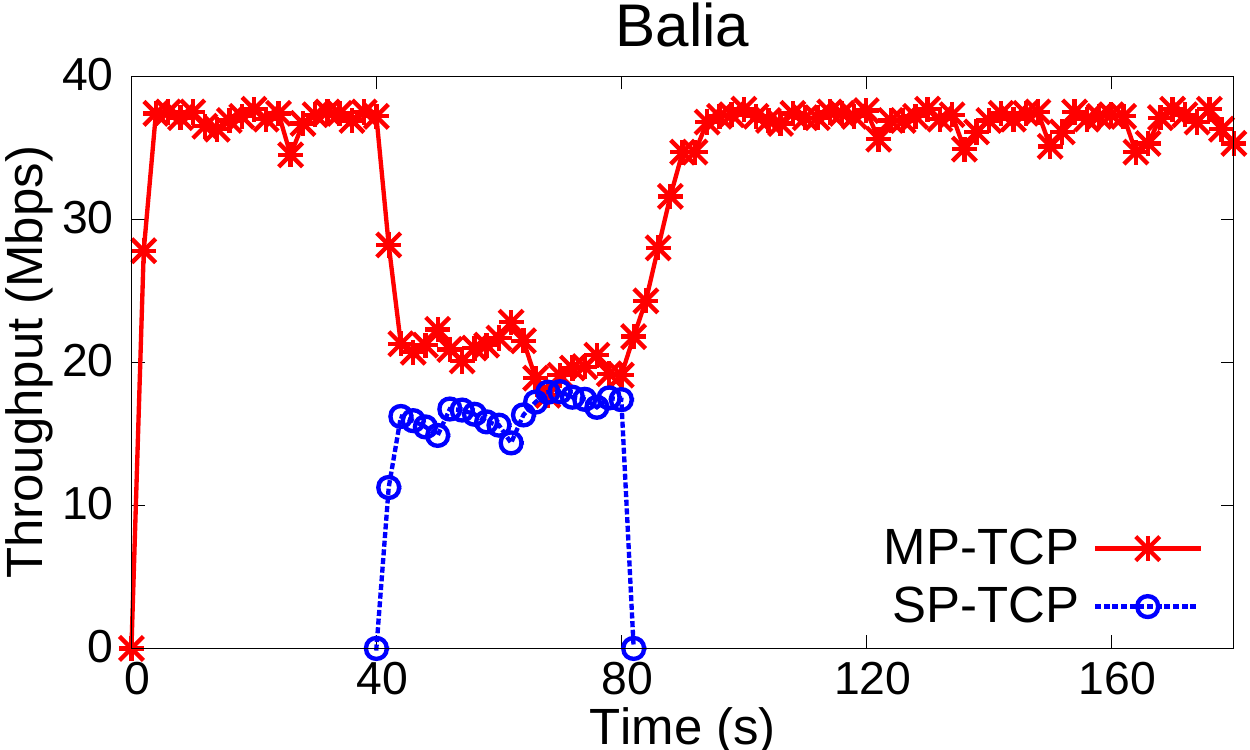}
\label{fig:exp_responsiveness_thru_ours}
	}
	\\
\subfloat{
	\includegraphics[width=0.45\linewidth,height=0.35\linewidth]{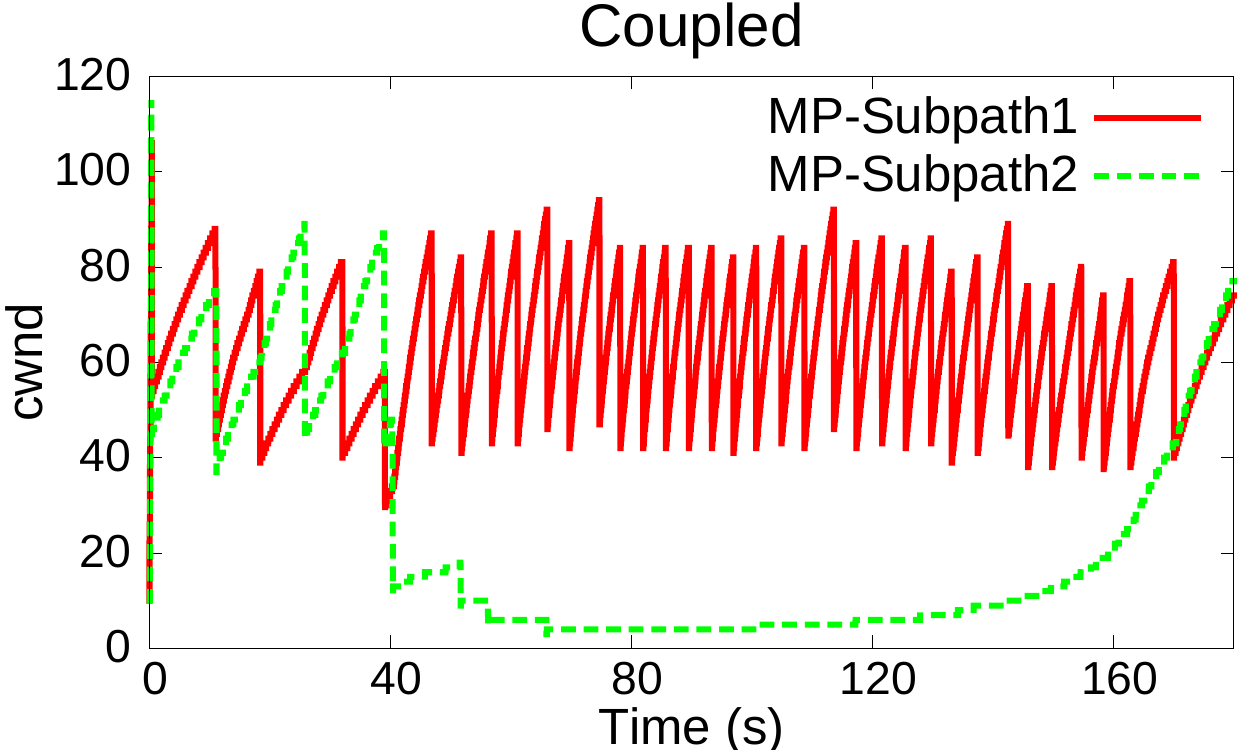}
\label{fig:exp_responsiveness_cwnd_coupled}
	}
\subfloat{
	\includegraphics[width=0.45\linewidth,height=0.35\linewidth]{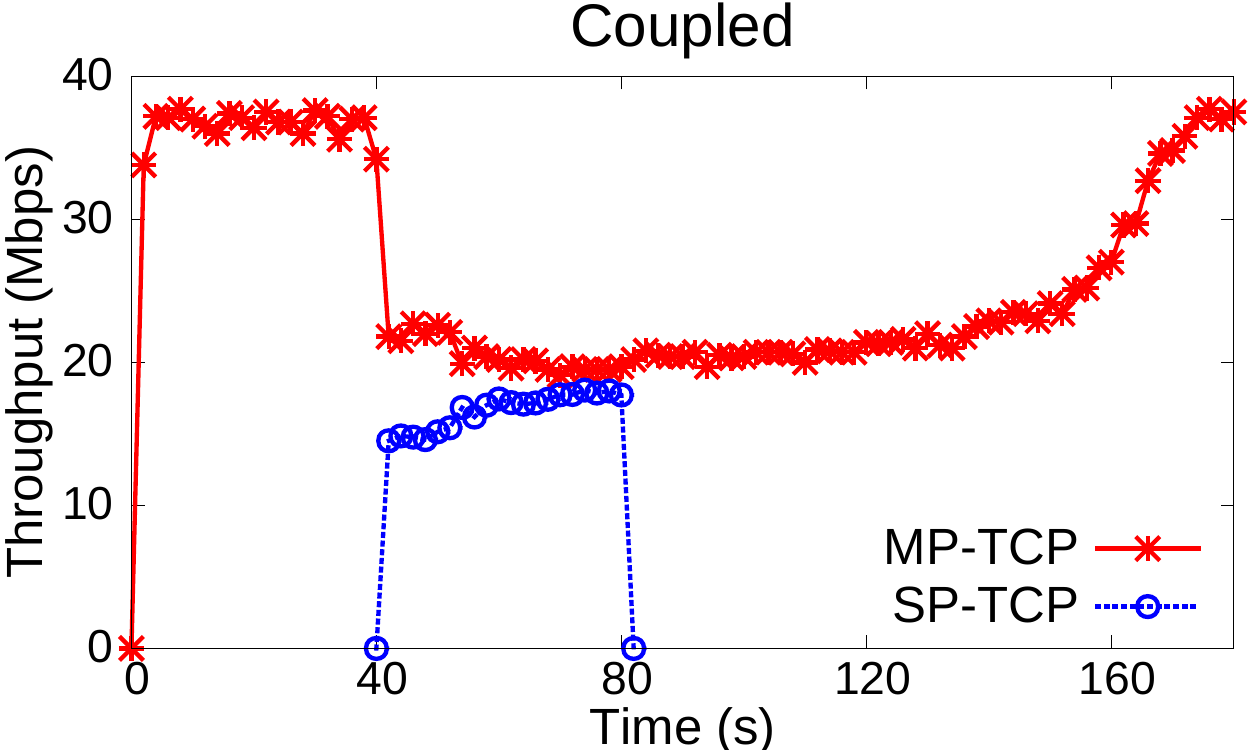}
\label{fig:exp_responsiveness_thru_coupled}
	}
	\\
\subfloat{
	\includegraphics[width=0.45\linewidth,height=0.35\linewidth]{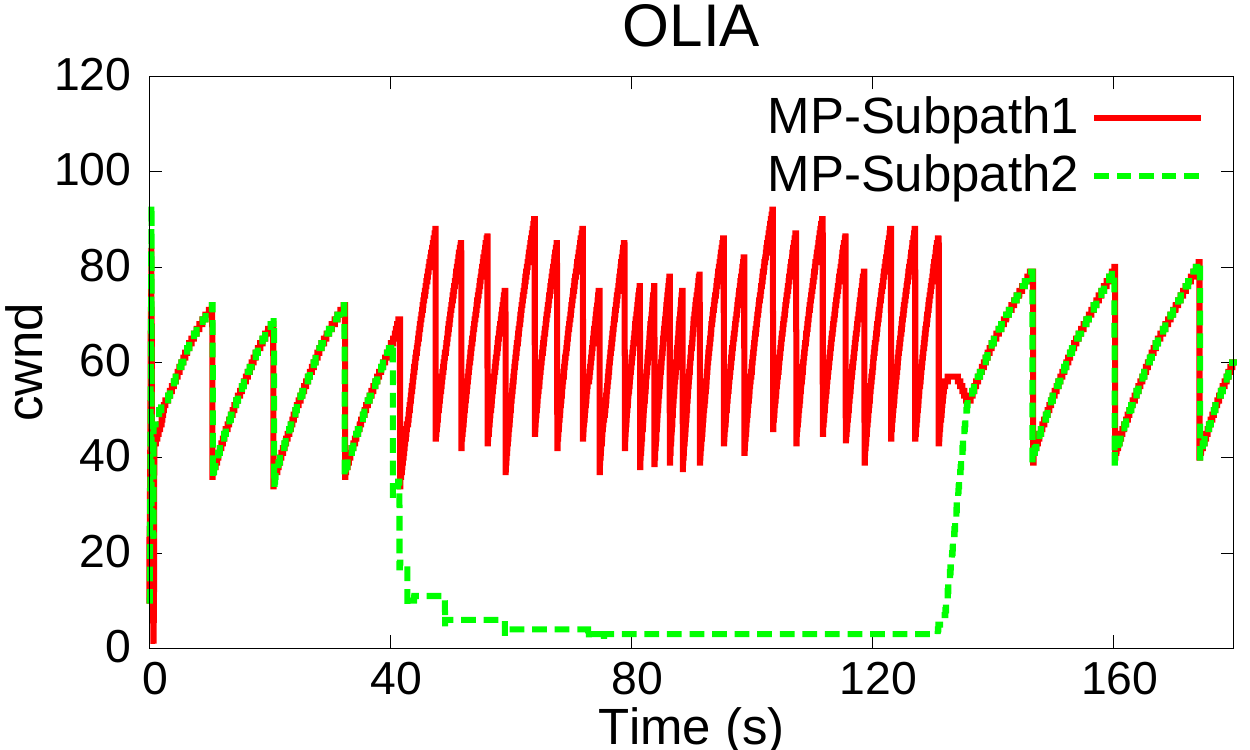}
\label{fig:exp_responsiveness_cwnd_olia}
	}
\subfloat{
	\includegraphics[width=0.45\linewidth,height=0.35\linewidth]{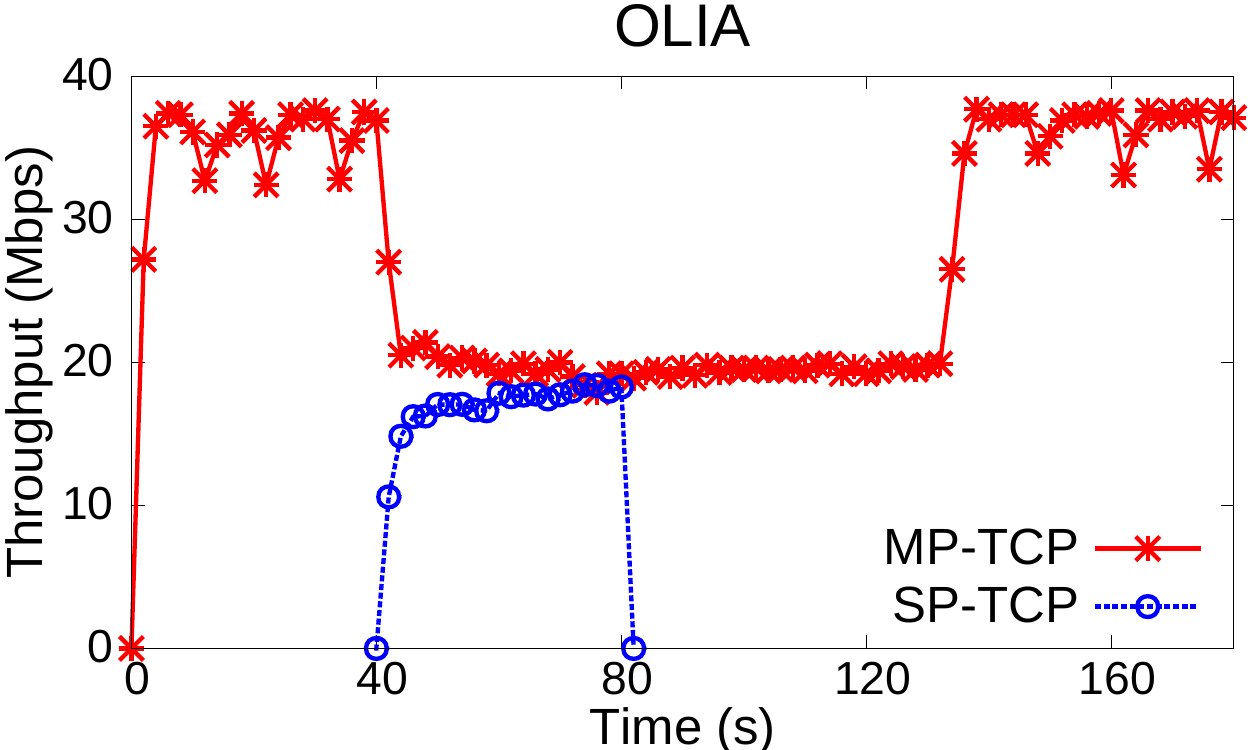}
\label{fig:exp_responsiveness_thru_olia}
	}
\caption{Responsiveness Performance: congestion window trajectory of MP-TCP for each path (left column). SP-TCP starts at time 40s and ends at 80s. The throughput of SP-TCP and total throughput of MP-TCP are shown in the right column. Parameters: $T_1=T_2=10$ms, $c_1=c_2=20$Mbps and $N_1=1, N_2=5$.}
\label{fig:dynamic}
\end{figure}

We use the network in Fig. \ref{fig:top1} with $c_1=c_2=20$Mbps, $T_1=T_2=10$ms
and $N_1=1,N_2=5$. To demonstrate the dynamic performance of each algorithm, we assume the MP-TCP flow is long lived while the single-path TCP flows
start at $40$s and end at $80$s. We record the aggregate throughput of the single-path TCP
flows from 40-80s, which measures the friendliness of MP-TCP.
We also measure the time for the congestion window on the second path to recover\footnote{Defined as
the first time the congestion window on the second path reaches the average congestion window
(e.g., $60$) after the single-path users have left.} of MP-TCP users.
It measures the responsiveness of MP-TCP. These measurements are shown in Table \ref{table:top1_d} and the congestion window and throughput trajectories of all algorithms are shown in Fig. \ref{fig:dynamic}. To clearly show the responsiveness performance, we record the longest convergence time found in our experiment in Table \ref{table:top1_d} and the corresponding trajectories are shown in Fig.~\ref{fig:dynamic}.

\begin{table}[htbp]
  \centering
  \caption{Responsiveness: convergence time (s) of MP-TCP and total throughput (Mbps) of all single-path TCP users.
  ($T_1=T_2=10$ms, $c_1=c_2=20$Mbps and $N_1=1, N_2=5$)}
  \begin{tabular}{|c|c|c|c|c|c|c|}
    \hline
             & ewtcp    & semi.    & max      & balia     & coupled  & olia         \\\hline
 Convergence & $3.25$   & $7.46$   & $17.75$  & $14.73$   & $94.36$  & $58.5$       \\\hline
 SP-TCP      & $13.89$  & $15.35$  & $15.8$   & $16.28$   & $16.64$  & $16.97$      \\\hline
  \end{tabular}
\label{table:top1_d}
\end{table}

EWTCP is the most responsive among all the algorithms.   Ours is as responsive as
the Max algorithm, yet significantly friendlier than EWTCP. Both Coupled and OLIA algorithms take an excessively long time to recover. For Coupled algorithm, the excessively slow recovery of the congestion window on the second path (see Fig. \ref{fig:dynamic}) is due to the design that increases the window roughly by $w_r/(\sum_{k\in s} w_k)^2$ on each ACK assuming the RTTs are similar.  After the single-path TCP flow has left, $w_2$ is small while $w_1$ is large, so that $w_2/(w_1+w_2)^2$ is very small.  It therefore takes a long time for $w_2$ to increase to its steady state value.  In general, under the Coupled algorithm, a route with a large throughput can greatly suppress the throughput on another route even though the other route is underutilized. The reason of the poor responsiveness performance of OLIA can be explained using similar argument as Coupled algorithm since they have the same increment/decrement for each ACK/loss in this scenario.

\subsection{Window oscillation}

\begin{figure}[htbp]
\centering
\begin{minipage}[b]{0.49\linewidth}
\includegraphics[width=\linewidth,height=0.7\linewidth]{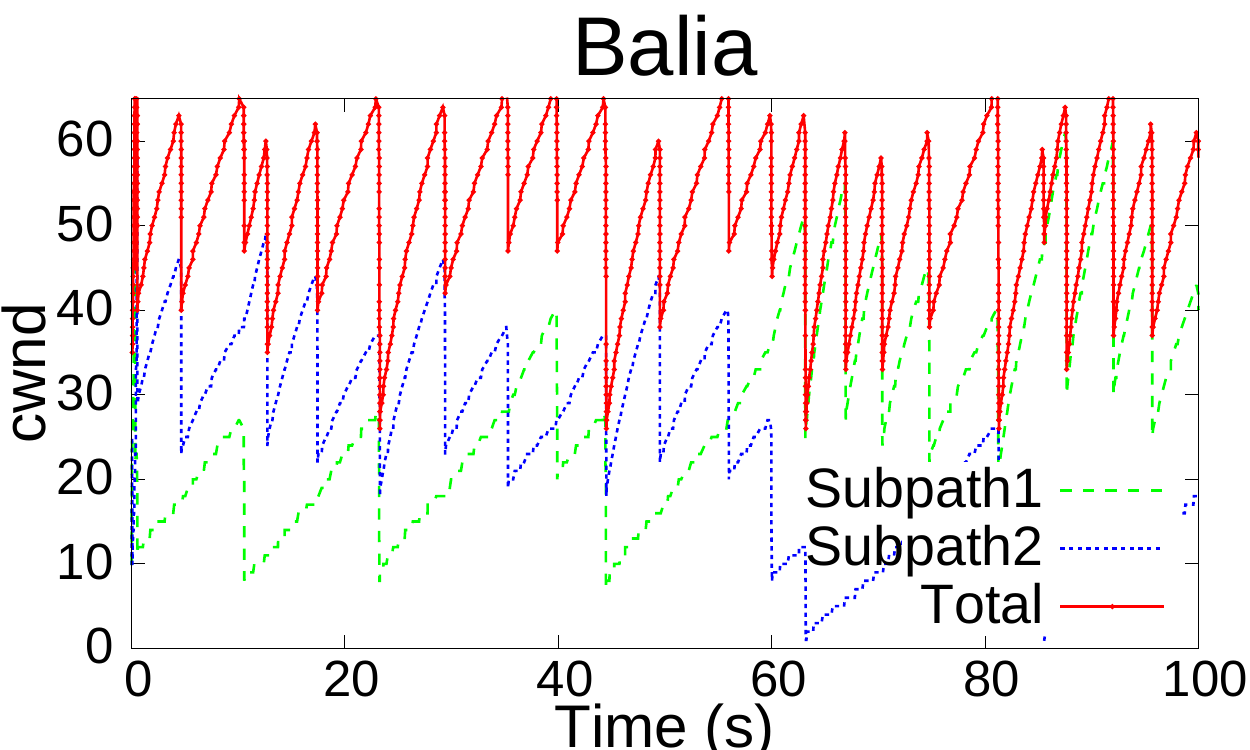}
\label{fig:exp_oscillation_ours}
\end{minipage}
\begin{minipage}[b]{0.49\linewidth}
\includegraphics[width=\linewidth,height=0.7\linewidth]{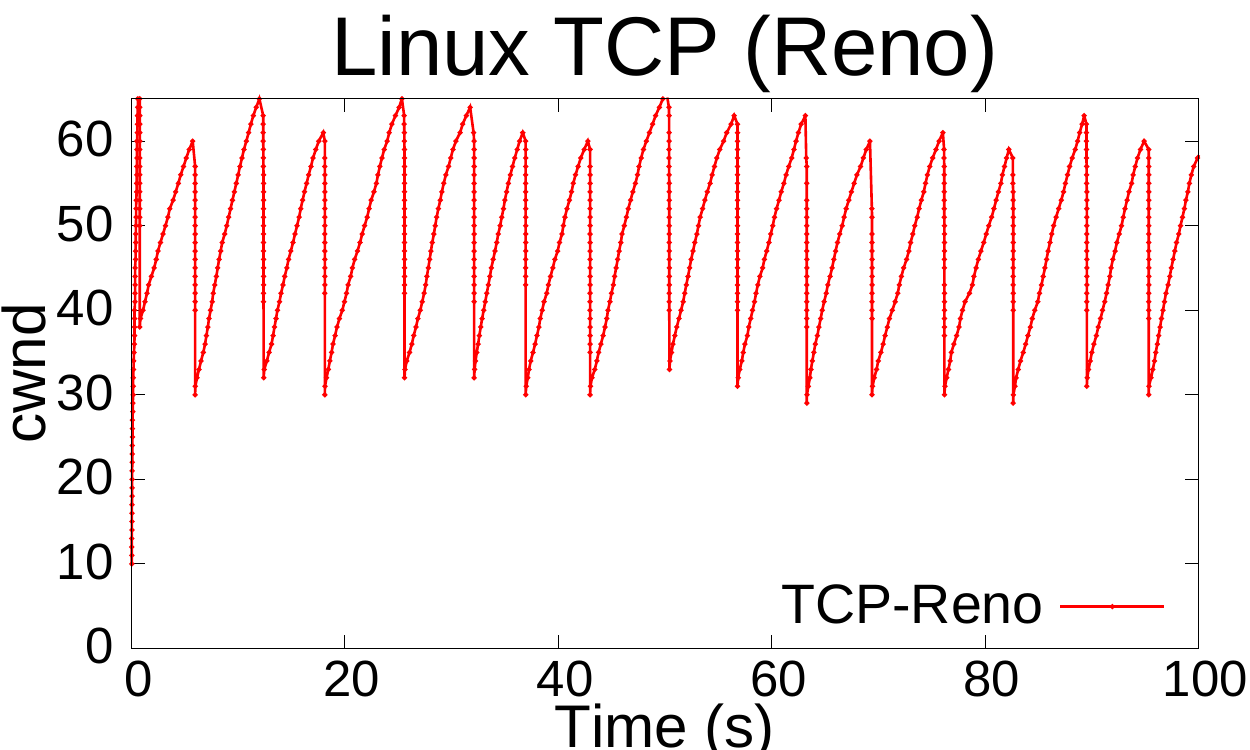}
\label{fig:exp_oscillation_reno}
\end{minipage}
\caption{Window oscillation: the red trajectories represent throughput fluctuations
	experienced by the application in the case of MP-TCP and the case of single-path TCP.}
\label{fig:packet_stability}
\end{figure}

We use a single-link network model to compare window oscillation under MP-TCP
and single-path TCP. First a MP-TCP flow initiates two subpaths through that link, and
we measure the window size of each subpath and their aggregate window size.
Then a TCP-Reno flow traverses the same link and we measure its window size.
 The results are shown in Fig. \ref{fig:packet_stability} for our algorithm in comparison
 with single-path TCP (other MP-TCP algorithms have a similar behavior).  They confirm
 that enabling multiple paths reduces the average window oscillation compared with only using single path.

\section{Conclusion}\label{sec:conc}

We have presented a model for MP-TCP and identified designs that guarantee the existence, uniqueness and stability of the network equilibrium.
We have characterized the design space and study the tradeoff
among TCP friendliness, responsiveness, and window oscillation.
We have proposed \emph{Balia } MP-TCP algorithm that generalizes existing
algorithms and strikes a good balance among these properties.
We have implemented \emph{Balia } in the Linux kernel and used it to evaluate
the performance of our algorithm.



\bibliographystyle{IEEEtran}
\bibliography{texcode/8_bibliography}{}

\section*{ACKNOWLEDGMENTS}
This work was supported by ARO MURI through grant W911NF-08-1-0233, NSF NetSE through grant CNS 0911041, Bell Labs, Alcatel-Lucent and Seoul R\&BD Program funded by the Seoul Metropolitan Government through grant WR080951.

\appendices

\section{Proof of Theorem \ref{thm:uf} (utility maximization)}
\label{app:uf}
The Lagrangian of \eqref{eqn:NUM} is:
\begin{align*}
L(\bx,\bfp)&=\sum_{s\in S}U_s( \bx_s)-\sum_{l\in L}p_l(y_l- c_l)\\
&=\sum_{s\in S}U_s( \bx_s)-\sum_{l\in L}p_l(\sum_{r\in R}H_{lr}x_r- c_l)\\
&=\sum_{s\in S}\left(U_s( \bx_s)-\sum_{r\in s}x_rq_r\right)+\sum_{l\in L}p_lc_l
\end{align*}
where $\bfp\geq \mathbf{0}$ are the dual variables and $q_r := \sum_{r\in R}H_{lr}p_l$.
Then the dual problem is
\bqn
D(\bfp)=\sum_{s\in S}\max_{\bx_s \geq 0}\{B_s(\bx_s,\bfp)\}+\sum_{l\in L}p_lc_l \quad \bfp\geq \mathbf{0}
\eqn
where $B_s( \bx_s, \bfp)=U_s(\bx_s)-\sum_{r\in s}x_r q_r$.
The KKT condition implies that, at optimality, we have
\begin{align}
&\frac{\partial U_s(\bx_s)}{\partial x_r}<q_r\Rightarrow x_r=0 \mbox{ and } x_r>0 \Rightarrow \frac{\partial U_s(\bx_s)}{\partial x_r}=q_r\label{eqn:utility1}\\
& y_l<c_l \Rightarrow p_l=0 \mbox{ and } p_l>0 \Rightarrow y_l=c_l\label{eqn:existence_p1}
\end{align}
Comparing with \eqref{eqn:existence_x}--\eqref{eqn:existence_p} we conclude that, if a MP-TCP
algorithm defined by (\ref{eqn:TCP_dynamics})--(\ref{eqn:AQM_dynamics}) has an underlying
utility function $U_s$, then we must have
\bq\label{eqn:equiv1}
\frac{\partial U_s(\bx_s)}{\partial x_r}= \phi_r(\bx_s) \quad r\in s, \ x_r>0
\eq
Given $\phi_r(\bx_s)$, \eqref{eqn:equiv1} has a continuously differentiable solutions $U_s(\bx_s)$
 if and only if the Jacobian of $\Phi_s(\bx_s)$ is symmetric, i.e., if and only if
\bqn
\frac{\partial \Phi(\bx_s)}{\partial \bx_s}=\left[\frac{\partial \Phi(\bx_s)}{\partial \bx_s}\right]^T
\eqn

\section{Proof of Theorem \ref{thm:existence} (existence and uniqueness)}
\label{app:existence}

\subsection{Proof of part 1}

For any link $l\in L$, let
$$\bfp_{-l}=\{p_1,\ldots,p_{l-1},p_{l+1},\ldots, p_{|L|}\},$$
whose component composes of all the elements in $\bfp$ except $p_l$. For $l\in L$, let
\begin{displaymath}
g_l(\bfp):=c_l-\sum_{r:l\in r}x_r=c_l-\sum_{s:r\in s, l\in r}y^s_l(p_l,\bfp_{-l})
\end{displaymath}
and $h_l(\bfp):=-g^2_l(\bfp)$. According to C1, we have the following two facts, which will be used in the proof.
\bi
\item $g_l(\bfp)$ is a nondecreasing function of $p_l$ on $\mathbb{R}_+$ since $y^s_l(\bfp)$ is a nonincreasing function of $p_l$.
 \item $\lim_{p_l\rightarrow \infty}g_l(p_l,\bfp_{-l})=c_l$ since $\lim_{p_l\rightarrow \infty}y^s_l(\bfp)=0$.
\ei

Next, we will show that $h_l(\bfp)$ is a quasi-concave function of $p_l$. In other words, for any fixed $\bfp_{-l}$, the set $S_a:=\{p_l\mid h_l(\bfp)\geq a\}$ is a convex set. If $g_l(0,\bfp_{-l})\geq 0$, then
\bqn
g_l(p_l,\bfp_{-l})\geq g_l(0,\bfp_{-l})\geq 0 \quad \forall p_l\geq 0,
\eqn
which means $h_l(p_l,\bfp_{-l})$ is a nonincreasing function of $p_l$, hence is a quasi-concave function of $p_l$ and
\bq\label{eqn:nash1}
\arg\max_{p_l}h_l(p_l,\bfp_{-l})=0.
\eq
On the other hand, if $g_l(0,\bfp_{-l})< 0$, then there exists a $p_l^*>0$ such that $g_l(p_l^*,\bfp_{-l})=0$ since $g_l(\cdot)$ is continuous and $\lim_{p_l\rightarrow \infty}g_l(p_l,\bfp_{-l})=c_l>0$. Note that $g_l(\bfp)$ is a nondecreasing function of $p_l$, then $h_l(p_l,\bfp_{-l})$ is nondecreasing for $p_l\in[0,p_l^*]$ and nonincreasing for $p_l\in[p_l^*,\infty)$. Hence,  $h_l(p_l,\bfp_{-l})$ is also a quasi-concave function of $p_l$ in this case and
\bq\label{eqn:nash2}
\max_{p_l}h_l(p_l,\bfp_{-l})=0.
\eq
By Nash theorem, if $h_l(p_l,\bfp_{-l})$ is a quasi-concave function of $p_l$ for all $l\in L$ and $\bfp$ is in a bounded set, then there exists a $\bfp^\star\in\mathbb{R}_+^{|L|}$ such that
\begin{displaymath}
p^\star_l=\arg\max_{p_l\in \mathbb{R}_+}h_l(p_l,\bfp_{-l}^*).
\end{displaymath}
According to \eqref{eqn:nash1} and \eqref{eqn:nash2}, for any $l\in L$, either $p_l^*>0$ or $g_l^*(\bfp^*)>0$ but not both holds at any time. Therefore $\bfp^*$ satisfies Eqn. \eqref{eqn:existence_p}. Since $\bfq=R^T\bfp$, there exists an $\bx^*$ to
\eqref{eqn:existence_x}.  Hence there exists at least one solution $(\bx,\bfp)$ that satisfies
\eqref{eqn:existence_x} and \eqref{eqn:existence_p}.

\subsection{Proof of part 2}


\begin{lemma}\label{contraction}
Assume a function $F: \mathbb{R}^{n}\rightarrow\mathbb{R}^{n}$ is continuously differentiable
and $\left[ \frac{\partial F} {\partial \bx} (\bx) \right]^+$ is negative definite for all $\bx$.  Then for any $\bx_1\neq\bx_2\in \mathbb{R}^{n}$,
\begin{displaymath}
(\bx_1-\bx_2)^T(F(\bx_1)-F(\bx_2))<0.
\end{displaymath}
\end{lemma}

\begin{IEEEproof}
Fix any $\bx_1 \neq \bx_2\in\mathbb{R}^n$.    Define $A(t) := F \left(t\bx_1+(1-t)\bx_2 \right)$.
Since $\partial F/\partial \bx$ is continuous, there exists a $\lambda< 0$ such that the eigenvalues
of $[\partial F / \partial \bx]^+ \leq \lambda$ over the compact set $\{t\bx_1+(1-t)\bx_2\mid 0\leq t\leq 1\}$.
Then
\bqn
&&(\bx_1-\bx_2)^T(F(\bx_1)-F(\bx_2))\\
&=&\int_0^1(\bx_1-\bx_2)^T\, \frac{d A}{d t}(\tau)\, d\tau\\
&=&\int_0^1(\bx_1-\bx_2)^T\, \frac{\partial F}{\partial \bx} \left(\tau \bx_1 + (1-\tau) \bx_2\right)\,
			(\bx_1-\bx_2)\ d\tau\\
&\leq &\lambda \|\bx_1-\bx_2\|_2^2<0
\eqn
\end{IEEEproof}

\begin{lemma}\label{lem:punique}
Suppose C3 holds. Then $x_r^*>0$ at equilibrium for all $r\in R$.
\end{lemma}

\begin{IEEEproof}
Suppose $x_r^* = 0$.  Then $q_r^* \geq \phi_r(\bx_r^*) = \infty$ by C3 and hence
there is a link $l \in r$ with $p_l^* = \infty$.   But then, for all paths $r' \ni l$, $q_{r'}^* = \infty$
and hence $x_{r'}^* = 0$ by C3.   This implies $y_l^* = 0 < c_l$, and hence $p_l^*=0$
by \eqref{eqn:existence_p}, contradicting $p_l^* = \infty$.
\end{IEEEproof}

Recall the vector notations that $\bx := (\bx_s, s\in S) := (x_r, r\in s, s\in S)$ and
$\Phi(\bx) := (\Phi_s(\bx_s), s\in S) := (\Phi_r(\bx_s), r\in s, s\in S)$.
To prove uniqueness of the equilibrium, suppose for the sake of contradiction
that there are two distinct equilibrium
points $(\bx, \bfp)$ and $(\hat{\bx}, \hat{\bfp})$.   By Lemma \ref{lem:punique} we have
$\bx>0$ and $\hat{\bx}>0$.  Hence \eqref{eqn:existence_x} implies $\Phi(\bx) = \bfq = H^T \bfp$
and $\Phi(\hat{\bx}) = \hat{\bfq} = H^T \hat{\bfp}$.
By Lemma \ref{contraction} and assumption C2 we then have
\bqn
0 & > & (\bx - \hat{\bx})^T (\Phi(\bx) - \Phi(\hat{\bx}) )
\\
& = & (\bx - \hat{\bx})^T H^T (\bfp - \hat{\bfp})
\\
& = & (\bfp - \hat{\bfp})^T (\by - \hat{\by})
\eqn
Hence
\bq
\bfp^T \by + \hat{\bfp}^T \hat{\by} & < &
\bfp^T \hat{\by} + \hat{\bfp}^T {\by}
\label{eq:100}
\eq
Equilibrium condition \eqref{eqn:existence_p} implies
\bq
\bfp^T (\bc - \by)  \ = \  0 & \text{and} & \hat{\bfp}^T (\bc - \hat{\by}) \ = \ 0
\label{eq:101}
\\
\by \ \leq \ \bc  & \text{and} & \hat{\by} \ \leq \ \bc
\label{eq:102}
\eq
Substituting \eqref{eq:101} into \eqref{eq:100} yields
\bqn
\bfp^T \bc + \hat{\bfp}^T {\bc} & < &
\bfp^T \hat{\by} + \hat{\bfp}^T {\by}
\\
\bfp^T (\bc - \hat{\by}) + \hat{\bfp}^T (\bc - \by) & < & 0
\eqn
But \eqref{eq:102} implies that the left-hand side of the last inequality is nonnegative
(since $\bfp \geq 0$, $\hat{\bfp} \geq 0$), a contradiction.
Hence the equilibrium is unique.


\section{Proof of Theorem \ref{thm:stability} (stability)}
\label{app:stability}

We will construct a Lyapunov function and use LaSalle's invariance principle \cite{Khalil1996}
to prove global asymptotic stability of the unique equilibrium point
 $(\bx^*, \bfp^*)$.
Define $\delta \bx:=\bx-\bx^\star$, $\delta \bfp:=\bfp-\bfp^\star$.  Consider the candidate
Lyapunov function:
\begin{equation}
\label{eq:lyapunov}
V(\bx,\bfp)=\sum_{r\in R}\int_{x_r^*}^{x_r}\frac{z-x_r^*}{k_r(z)}dz +
	\frac{1}{2}\sum_{l\in L}\frac{\delta p_l^2}{\gamma_l}
\end{equation}
By definition, $V(\bx,\bfp) > 0$ for all $(\bx,\bfp) \neq (\bx^*,\bfp^*)$  and
$V(\bx,\bfp) = 0$ if $(\bx,\bfp) = (\bx^*,\bfp^*)$.
Furthermore $V$ is radially unbounded, i.e.,
 $V(\bx, \bfp)\rightarrow \infty$ as $\|(\bx, \bfp)\|_2 \rightarrow \infty$.
Finally
\bqn
\dot{V}(\bx,\bfp)=\sum_{r\in R}\frac{1}{k_r(x_r)}\delta x_r\dot{x}_r+\sum_{l\in L}\frac{1}{\gamma_l}\delta p_l\dot{p}_l
\eqn
If $\delta x_r \neq 0$ then we have (since $k_r(\bx_s) = k_r(x_r)$)
\bqn
\frac{1}{k_r(x_r)}\delta x_r\dot{x}_r &=&\delta x_r\left(\phi_r(\bx_s)-q_r\right)_{x_r}^+\\
&\leq&\delta x_r\left(\phi_r(\bx_s)-q_r\right)\\
&=& \delta x_r\left(\phi_r(\bx_s)-\phi_r(\bx_s^*)- \delta q_r\right)
\eqn
The first inequality holds since $(\phi_r(\bx_s)-q_r)_{x_r}^+=\phi_r(\bx_s)-q_r$ if $x_r>0$ and $\phi_r(\bx_s)-q_r\leq 0$, $\delta x_r=-x_r^*$ if $x_r=0$.
The last equality holds since $\phi_r(\bx_s^*) = q_r^*$ by Lemma \ref{lem:punique}
and \eqref{eqn:existence_x}.    Hence
\bqn
\sum_{r\in R} \frac{1}{k_r(x_r)}\delta x_r\dot{x}_r & \leq &
\delta \bx^T (\Phi(\bx) - \Phi(\bx^*)) - \delta \bx^T \delta \bfq
\\
&  < & - \delta \bx^T H^T \delta \bfp
\eqn
where the last inequality holds since  $\delta \bx^T \left(\phi(\bx)-\phi(\bx^*)\right) < 0$
by  Lemma \ref{contraction} and assumption C2.
Similarly
\bqn
\frac{1}{\gamma_l}\delta p_l\dot{p}_l \ = \ \delta p_l(y_l-c_l)_{p_{l}}^{+}
	\ \leq \ \delta p_l(y_{l}-c_l)
	\ \leq \ \delta p_l\delta y_l
\eqn
where the last inequality holds since $\delta p_l c_l \geq \delta p_l y_l^*$ by the equilibrium
condition \eqref{eqn:existence_p}.
Hence
\bqn
\sum_{l\in L} \frac{1}{\gamma_l}\delta p_l\dot{p}_l  & \leq & \delta \bfp^TH \delta\bx
\eqn
Therefore if $\delta \bx\neq 0$ then
\bqn
\dot{V}(\bx,\bfp)&<& -\delta \bx^T H^T \delta\bfp+\delta \bfp^TH \delta\bx \ = \ 0
\eqn
and if $\delta\bx =0$ then $\dot{V}(\bx, \bfp) = 0$.
This means $\dot{V}(\bx, \bfp) \leq 0$ and $V$ is indeed a Lyapunov function.

Consider the set
\bqn
Z  & := & \{ \ (\bx(t), \bfp(t)) \, | \, \dot{V}(\bx(t), \bfp(t)) = 0  \text{ for all } t\geq 0 \ \}
\eqn
of trajectories on which $\dot{V} \equiv 0$.
If the only trajectory in $Z$ is the trivial trajectory $(\bx, \bfp) \equiv (\bx^*, \bfp^*)$ then
LaSalle's invariance principle  implies that $(\bx^*, \bfp^*)$ is globally asymptotically stable.
We now show that this is indeed the case.

As shown above $\dot{V}  \equiv 0$ implies $\delta \bx \equiv 0$, i.e., any trajectory
$(\bx(t), \bfp(t))$ in $Z$ must have $\bx(t) = \bx^*$ for all $t\geq 0$.  This means
$\dot{\bx} \equiv 0$ and hence, for all $t\geq 0$,
$\bfq(t) = \Phi(\bx(t))$ since $\bx(t) = \bx^*>0$ by Lemma \ref{lem:punique}.
That is, for all $t\geq 0$, $H^T \bfp(t) = \Phi(\bx^*)$  and hence
$\bfp(t) = \bfp^*$ since $H$ has full row rank by C3.
Therefore  $(\bx, \bfp) \equiv (\bx^*, \bfp^*)$ is indeed the only trajectory in $Z$.
This completes the proof of Theorem \ref{thm:stability}.

\section{Proof of Theorem \ref{thm:friendliness} (friendliness)}
\label{app:definition_fairness}

Let the MP-TCP source be defined by
\bqn
\phi_r(\bx_s;\mu)=\mu\tilde\phi_r(\bx_s)+(1-\mu)\hat{\phi}_r(\bx_s), \quad \mu\in [0,1]
\eqn
Algorithm $\hat M$ and $\tilde M$ corresponds to $\mu=0$ and $\mu=1$ respectively.
Let $x_g$ and $\tau_g$ be the throughput and RTT of the TCP NewReno source in
Fig.~\ref{fig:fairness_illustration}.
The equilibrium is defined by $F(\bx, \mu) = 0$ where $\bx := (\bx_s, x_g)$ and $F$ is given by:
\bqn
\Phi_s(\bx_s; \mu) - \frac{1}{\tau_g^2x_g^2} \mathbf{1} & = & 0\\
\mathbf{1}^T \bx_s  +  x_g  & = & c
\eqn
where the first equation follows from
\bqn
p^*=\frac{1}{\tau_g^2x_g^2}=\phi_r(\bx_s; \mu), \quad r\in s
\eqn
and $p^*$ is the congestion price at the bottleneck link. Applying
the implicit function theorem, we get
\bqn
\frac{d \bx}{d\mu}  &=& - \left( \frac{\partial F}{\partial \bx} \right)^{-1} \frac{\partial F}{\partial \mu}\\
 &=& -  \begin{bmatrix}
		\frac{\partial \Phi_s}{\partial \bx_s}  &  \frac{2}{x_g^3}\, \mathbf{1}
		\\
		\mathbf{1}^T  &  1
	\end{bmatrix}^{-1}
	\begin{bmatrix}
	\tilde{\Phi}_s(\bx_s) - \hat{\Phi}_s(\bx_s)   \\   0
	\end{bmatrix}
\eqn
where the inverse exists by condition C2.
C2 also guarantees the inverse of $\frac{\partial \Phi_s}{\partial\bx_s} (\bx_s; \mu)$,
denoted by $D(\mu)$;
C4 ensures $\sum_{i\in s}D_{ij}(\mu)\leq 0$.
Let
\bqn
A \ := \ \frac{\partial \Phi_s}{\partial \bx_s} - \frac{2}{x_g^3} \mathbf{1}\mathbf{1}^T
& \mbox{ and } &
d \ := \ 1 - \frac{2}{x_g^3} \sum_{i, j} D_{ij}(\mu)
\eqn
Then
\bqn
 \begin{bmatrix}
		\frac{\partial \Phi_s}{\partial \bx_s}  &  \frac{2}{x_p^3}\, \mathbf{1}
		\\
		\mathbf{1}^T  &  1
	\end{bmatrix}^{-1}
& = &
\begin{bmatrix}
A^{-1}  &  -D \mathbf{1} d
\\
- d \mathbf{1}^T A^{-1}  &  d^{-1}
\end{bmatrix}
\eqn
Thus
\bq\label{eqn:artificial1}
\mathbf{1}^T\frac{\partial \bx_s}{\partial \mu}&=&-[\mathbf{1}^T 0] \left( \frac{\partial F}{\partial \bx} \right)^{-1} \frac{\partial F}{\partial \mu}\nonumber\\
&=&-\mathbf{1}^TA^{-1}(\tilde{\Phi}_s(\bx_s) - \hat{\Phi}_s(\bx_s))
\eq
By matrix inverse formula,
\bqn
A^{-1}&=&\left( \frac{\partial \Phi_s}{\partial \bx_s} - \frac{2}{x_g^3} \mathbf{1}\mathbf{1}^T\right)^{-1}\\
&=&D(\mu)+\frac{1}{\frac{x_g^3}{2}-\mathbf{1}^TD(\mu)\mathbf{1}}D(\mu)\mathbf{1}\mathbf{1}^TD(\mu)
\eqn
Substitute it into \eqref{eqn:artificial1}, we have
\begin{align*}
&\mathbf{1}^TA^{-1}(\hat{\Phi}_s(\bx_s) - \tilde{\Phi}_s(\bx_s))\\
=&\left(1+\frac{\mathbf{1}^TD(\mu)\mathbf{1}}{\frac{x_g^3}{2}-\mathbf{1}^TD(\mu)\mathbf{1}}\right)\mathbf{1}^TD(\mu)(\tilde{\Phi}_s(\bx_s) - \hat{\Phi}_s(\bx_s))\\
=&\frac{x_g^3}{x_g^3-2\mathbf{1}^TD(\mu)\mathbf{1}}\sum_{r\in s}\left(\sum_{i\in s}D_{ir}(\mu)\right)(\tilde{\phi}_r(\bx_s) - \hat{\phi}_r(\bx_s))\\
\leq& 0
\end{align*}
where the inequality follows because $D(\mu)$ is negative definite,
$\sum_{i\in s}D_{ir}(\mu)<0$ and $\tilde{\phi}_r(\bx_s) - \hat{\phi}_r(\bx_s)\geq 0$.
Thus we have $\mathbf{1}^T\frac{\partial \bx_s}{\partial \mu}\geq0$ for $\mu\in[0,1]$,
i.e., the aggregate throughput of the MP-TCP over its available paths is increasing
in $\mu$.   This means $\tilde M$ $($corresponding to $\mu=1)$ will attain a higher
throughput than $\hat M$ $($corresponding to $\mu=0)$
when separately sharing the test network in Fig.~\ref{fig:fairness_illustration}
 with the same SP-TCP.


\section{Proof of Theorem \ref{thm:SR} (responsiveness)}

\subsection{Proof of part 1}

Fix any eigenvalue $\lambda$ of $J^*$.
Let $\bz := (\bx, \bfp) \in Z$ be the corresponding eigenvector with $\|\bz\|_2 = 1$.
Then we have
\bqn
\lambda \begin{bmatrix}
\bx \\
\bfp
\end{bmatrix}=
\begin{bmatrix}
\Lambda_{k}& 0\\
 &  \Lambda_{\gamma}\\
\end{bmatrix}
\begin{bmatrix}
\frac{\partial \Phi}{\partial \bx} & -H^T\\
H & 0\\
\end{bmatrix}
\begin{bmatrix}
\bx \\
\bfp
\end{bmatrix}
\eqn
Hence
\bqn
\lambda
\begin{bmatrix}
\Lambda_{k}^{-1}& 0\\
 &  \Lambda^{-1}_{\gamma}\\
\end{bmatrix}
\begin{bmatrix}
\bx \\
\bfp
\end{bmatrix}=
\begin{bmatrix}
\frac{\partial \Phi}{\partial \bx} & -H^T\\
H & 0\\
\end{bmatrix}
\begin{bmatrix}
\bx \\
\bfp
\end{bmatrix}
\eqn
Premultiplying $\bz^H$ on both sides, we have
\bqn
\lambda &=&  \frac{ \bx^H \frac{\partial \Phi}{\partial \bx}\, \bx \ + \ (\bfp^H H \bx - \bx^H H^T \bfp) }
	{ \bx^H \Lambda_{k}^{-1} \bx + \bfp^H \Lambda_{\gamma}^{-1} \bfp }
\eqn
The denominator is real and positive, and $(\bfp^H H \bx - \bx^H H^T \bfp)$ in the
numerator is imaginary.  Hence
\bqn
\mathbf{Re} (\lambda) & = &
	\frac{ \mathbf{Re} \left( \bx^H \frac{\partial \Phi}{\partial \bx}\, \bx \right) }
	{ \bx^H \Lambda_{k}^{-1} \bx + \bfp^H \Lambda_{\gamma}^{-1} \bfp }
\\
& = &
	\frac{ \bx^H \left[ \frac{\partial \Phi}{\partial \bx} \right]^+ \bx  }
	{ \bx^H \Lambda_{k}^{-1} \bx + \bfp^H \Lambda_{\gamma}^{-1} \bfp }
	\ < \ 0
\eqn
where the last inequality holds because the numerator is negative by
condition C2 and the denominator is positive.
Since this holds for all eigenvalues $\lambda$ of $J^*$,
the linearized system \eqref{eq:ls} is stable.
Moreover $\mathbf{Re} (\lambda) \leq \overline{\lambda}(J^*) \leq  0$ as desired.

\subsection{Proof of part 2}

Consider two MP-TCP algorithms $(\hat K, \hat \Phi)$ and $(\tilde K, \tilde \Phi)$
such that
\bqn
\hat K_s \ \geq \ \tilde K_s & \text{and} &
\frac{\partial \hat\Phi_s}{\partial \bx_s}\preceq \frac{\partial \tilde\Phi_s}{\partial \bx_s}
\qquad \text{for all } s\in S
\eqn
For any (nonzero) $\bz = (\bx, \bfp) \in Z$ we have
\bq
0 \ \leq \ \bx^H\hat\Lambda_{k}^{-1} \bx  &\leq&
	\bx^H \tilde\Lambda_{k}^{-1}  \bx
\label{eq:1}
\\
	\bx^H \left[ \frac{\partial \hat\Phi}{\partial \bx} \right]^+ \bx
& \leq &  \bx^H \left[ \frac{\partial \tilde\Phi}{\partial \bx} \right]^+ \bx
\ < \ 0
\label{eq:2}
\eq
Hence $\overline\lambda(\hat J^*)  \leq \overline\lambda (\tilde J^*)$.


\section{Proof of Theorem \ref{thm:tradeoff} (tradeoff)}
\label{app:tradeoff}

Fix an $s$.
Let $f_r(\bx_s):=\hat\phi_r(\bx_s)-\tilde{\phi}_r(\bx_s)$ and
$F(\bx_s):=(f_r(\bx_s), r\in s) = \hat\Phi_s(\bx_s)-\tilde{\Phi}_s(\bx_s)$.
Suppose for the sake of contradiction that $\partial\hat\Phi_s(\bx_s)/\partial \bx_s \preceq \partial\tilde{\Phi}_s(\bx_s)/\partial \bx_s$
but $\hat\Phi_s(\bx_s) \geq \tilde\Phi_s(\bx_s)$ does not hold,
i.e., there exists a finite $\bx_s^0$ and a $r \in s$ such that
\bq
f_r(\bx^0_s)  \, = \, \hat\phi_r (\bx^0_s) - \tilde\phi_r (\bx^0_s) & < & 0
\label{eq:103}
\eq

Since $\left[ \partial F/\partial \bx_s \right]^+ \preceq 0$ by assumption, a trivial modification of
Lemma \ref{contraction} shows that, for all $\bx_s \neq \bx_s^0$,
$(\bx_s - \bx_s^0)^T (F(\bx_s) - F(\bx_s^0)) \leq 0$, i.e.,
\bq
0 & \geq & \sum_{r' \in s}\, (x_{r'}  - x^0_{r'} )\, (f_{r'} (\bx_s) - f_{r'} (\bx^0_s))
\label{eq:104}
\eq
Choose an $\bx_s$ as follows:
for all $r' \neq r$, choose $x_{r'} = x^0_{r'}$ and then use condition C5 to
choose an $x_r < \infty$ large enough so that  $x_r > x^0_r$ and $f_r(\bx_s) > f_r(\bx^0_s)/2$.
With this $\bx_s$, \eqref{eq:104} becomes
\bqn
0 & \geq &  (x_r  - x^0_r)\, (f_r (\bx_s) - f_r (\bx^0_s))
\\
& > &  (x_r - x^0_r)\, \left( - \frac{f_r(\bx^0_s)}{2} \right)  \ > \ 0
\eqn
where the last inequality follows from \eqref{eq:103}.  This is a contradiction and hence
$\hat\Phi_s(\bx_s) \geq \tilde\Phi_s(\bx_s)$.

\section{Proof of Theorem \ref{thm:propose}}\label{app:propose}

We will show the results hold for any $n\in\mathbb{N}_+$. Since $\lim_{n\rightarrow \infty}\|\bx_s\|_n=\|\bx_s\|_\infty$, the results also hold for $n=\infty$.
When $\beta=0$, it is easy to show that $\phi_r$ satisfies C1 and
$\left[ \frac{\partial \Phi_s}{\partial \bx_s} \right]^+$ is negative semidefinite
under the conditions of the theorem.  We hence prove the theorem for $\beta >0$.

\subsection{Proof of part 1}

Fix any $n\in \mathbb N_+$ and $\beta > 0$.
Fix any finite $\bfp\geq 0$ such that $q_r>0$ for all $r$.   Fix any $s\in S$.
We now show that there exists an $\bx_s > 0$ that satisfies \eqref{eqn:existence_x},
in particular $\phi_r(\bx_s) = q_r$,
in two steps.

First, there exists an $\bx_s$ that satisfies $\phi_r(\bx_s) = q_r$ if and only if
\bq
\phi_r(\bx_s) = \frac{2} { \tau_r^2\, \|\bx_s\|_1^{2} }  \left( 1 \ + \ \beta \left(
			\frac{ \|\bx_s\|_{n} } {x_r} - 1
			\right)
			\right)=q_r,
\label{eq:105}
\eq
which is equivalent to

\bq
\frac{x_r}{\|\bx_s\|_n} & = & \frac{2\beta}{ 2\beta + q_r \tau_r^2 \|\bx_s\|_1^2 - 2 }
\label{eq:106}
\eq
Since this holds for all $r\in s$, we have
\bq\label{eq:1106}
1 & = & \sum_{r\in s} \left( \frac{x_r}{\|\bx_s\|_n} \right)^n
\\
& = & \sum_{r\in s} \left( \frac{2\beta}{  2\beta + q_r \tau_r^2 \|\bx_s\|_1^2 - 2 } \right)^n \ =: \ \psi \left( \|\bx_s\|_1^2 \right)\nonumber
\eq
Clearly $\psi\left( C \right) \rightarrow 0$ as $C \rightarrow \infty$.
Let
\bq
\underline{C} & := & \frac{2}{\min_{r\in s} \, q_r \tau_r^2}
\label{eq:107}
\eq
Then $\underline{C} < \infty$ since $q_r>0$ for all $r$ by assumption.   Moreover
 $q_r \tau_r^2 \underline{C} \geq 2$ for all $r \in s$ and hence
\bqn
\psi\left( \underline{C} \right) & = & 1 +
		\sum_{r\neq \underline{r}} \left( \frac{2\beta}{  2\beta + q_r \tau_r^2 \underline{C} - 2 } \right)^n
		\ > \ 1
\eqn
where $\underline{r}$ is a minimizing $r\in s$ in \eqref{eq:107}.
Since $\psi(C)$ is continuous, there exists an $\tilde{C} \in [\underline{C}, \infty)$ with
$\psi (\tilde C) = 1$.
Moreover such a $\tilde C$ is unique since $\psi(C)$ is strictly decreasing.

Finally consider the set of $\bx_s$ with $\|\bx_s\|_1^2 = \tilde C$.  All such $\bx_s$ satisfy
\eqref{eq:106} with
\bq
x_r & = & \frac{2\beta}{  2\beta + q_r \tau_r^2 \tilde C - 2 }\, \|\bx_s\|_n \ =: \ a_r \, \|\bx_s\|_n
\label{eq:108}
\eq
But $\tilde C = \|\bx_s\|_1^2 = \left( \sum_{r\in s} a_r\, \|\bx_s\|_n \right)^2$, implying
\bqn
\|\bx_s\|_n & = & \frac{\sqrt{\tilde C}} { \sum_{r\in s} a_r }
\eqn
In summary, given any finite $\bfp\geq 0$ such that $q_r>0$ for all $r$,
 a solution $\bx_s > 0$ to \eqref{eq:106} is \emph{uniquely}
given by
\bq
x_r & = & \frac{ a_r } { \sum_{k\in s} a_{k} } \sqrt{\tilde C}, \qquad r\in s
\label{eq:109}
\eq
where
\bqn
a_r & := & \frac{2\beta}{  2\beta + q_r \tau_r^2 \tilde C - 2 }
\eqn
and $\tilde C = \|\bx_s\|_1^2$ is the unique value at which $\psi(\tilde C) = 1$.

We now prove the other conditions in C1:
\bqn
\frac{\partial y^s_l(\bfp)}{\partial p_l}\leq0, \quad \quad \lim_{p_l\rightarrow \infty} y^s_l (\bfp)=0
\eqn
According to \eqref{eq:1106}, we can show that $\tilde C$ is a decreasing function of $q_r$ and $q_r\tau_r^2\tilde C$ is an increasing function of $q_r$ for $r\in s$. Thus, $\tilde C$ is a decreasing function of $p_l$ and $q_r\tau_r^2\tilde C$ is an increasing of $p_l$ if $l\in r$ because $q_r=\sum_{l\in L}H_{lr}p_l$. For each $l\in L$, let $s_l:=\{r\mid l\in r,r\in s\}$, then by definition and \eqref{eq:109}, we have
\bqn
y^s_l(\bfp)=\frac{\sum_{r\in s_l}a_r}{\sum_{r\in s}a_r}\sqrt{\tilde C}
=\frac{\sum_{r\in s_l}a_r}{\sum_{r\in s_l}a_r+\sum_{r\not\in s_l}a_r}\sqrt{\tilde C}.
\eqn
Since $a_r$ is a decreasing function of $q_r\tau_r^2\tilde C$, it is also a decreasing function of $p_l$ if $l\in r$. Recall that $\sqrt{\tilde C}$  is also a decreasing function of $p_l$, $y^s_l(\bfp)$ is thus a decreasing function of $p_l$, in other words, $\frac{\partial y^s_l(\bfp)}{\partial p_l}\leq0$.

On the other hand, as $p_l \rightarrow \infty$, $q_r \rightarrow \infty$ for all paths $r$ traversing $l$. Then $x_r\rightarrow 0$ by \eqref{eq:105} for $l\in r$, which shows $ \lim_{p_l\rightarrow \infty} y^s_l (\bfp)=0$.

\subsection{Proof of part 2}

To prove $\phi_r(\bx_s)$ satisfies C2 and C3 for $\beta>0$, we will show that the
Jacobian $\partial\Phi_s(\bx_s)/\partial \bx_s$ is negative definite if $0 < \beta \leq 1$,
$|s| \leq 8$ and $\tau_r$ are the same for $r\in s$.
Other properties of C2 and C3 are easy to prove and we omit the proof.
Fix an $s$ and let $\tau_r = \tau$, the common round-trip time for all $r\in s$.

Let $\Lambda_s := \text{diag}\{\bx_s\}$ and
\bqn
\mathbf{a}_s & := & \left(\frac{2x_r}{\|\bx_s\|_1}-\frac{x_r^{n}}{\|\bx_s\|_n^n}, \ r\in s\right)
\eqn
Then the Jacobian of $\Phi_s$ at $\bx_s$ is
\bqn
\frac{\partial \Phi_s}{\partial \bx_s}=-\frac{4(1-\beta)}{\tau^2\|\bx_s\|_1^3}\mathbf{1}\mathbf{1}^T-2\beta\frac{\|\bx_s\|_n}{\tau^2\|\bx_s\|_1^2}\Lambda_s^{-1}\left(I_{|s|}+\mathbf{1}\mathbf{a}_s^T\right)\Lambda_s^{-1}
\eqn
and it is negative definite for $\beta>0$ if $\left[ I_{|s|}+\mathbf{1}\mathbf{a}_s^T \right]^+$
is positive definite.   We now show that this is indeed the case when $|s| \leq 8$, i.e.,
for any $\mathbf{z}_s\in\mathbb{R}^{|s|}$,
\bq\label{eqn:psd1}
\mathbf{z}_s^T(I_{|s|}+\mathbf{1}\mathbf{a}_s^T)\mathbf{z}_s=
\|\mathbf{z}_s\|_2^2+ \sum_{r\in s}z_r  \sum_{r\in s}a_rz_r
\ > \ 0
\eq
By Lemma \ref{lem:sum2} below, $\mathbf{1}^T \mathbf{a}_s = 1$ and
$\|\mathbf{a}_s\|_2^2\leq 1$.
Then \eqref{eqn:psd1} follows from Lemma \ref{lem:psd1} below provided $|s|\leq 8$.
Hence the Jacobian is negative definite.\footnote{If $\beta=0$ the Jacobian degenerates
to
\begin{equation}
\frac{\partial \Phi_s}{\partial \bx_s}=-\frac{4}{\tau^2\|\bx_s\|_1^3}\mathbf{1}\mathbf{1}^T,
\end{equation}
which is merely negative semidefinite.
}
The proof of Theorem  \ref{thm:propose} is complete after Lemmas \ref{lem:sum2}
and \ref{lem:psd1} are proved.

To show that it satisfies C3, it follows directly from \eqref{eq:105} that if $x_r=0$ then $\phi_r(\bx_s)=\infty$.  It is also clear from \eqref{eq:105} that
the converse holds.  This proves C3.

\begin{lemma}\label{lem:sum2}
Fix  any integer $p \geq 1$.
Given any ${\bx} \in \mathbb{R}^m_+$, define a vector ${\mathbf{a}}$ in $\mathbb{R}^m$ as follows:
\bqn
a_i & = & \frac{2x_i}{\sum_{j=1}^m x_j}-\frac{x_i^{p}}{\sum_{j=1}^m x_j^p}, \quad 1\leq i\leq m
\eqn
Then $\sum_{i=1}^m a_i=1$ and $\sum_{i=1}^m a_i^2\leq 1$.
\end{lemma}
\begin{IEEEproof}
It is obvious that $\sum_{i=1}^m a_i=1$.    To show $\sum_{i=1}^m a_i^2\leq 1$, we have
\begin{align*}
\sum_{i=1}^m a_i^2&=\frac{\sum_i x_i^{2p}}{\left(\sum_j x_j^p\right)^2}
		+ \frac{4\sum_i x_i^2}{\left(\sum_j x_j\right)^2}
		- \frac{4\sum_i x_i^{p+1}}{\left(\sum_j x_j^p\right)\left(\sum_j x_j\right)}
\\
&\leq 1 + \frac{4\sum_i x_i^2}{\left( \sum_j x_j \right)^2}
		- \frac{4\sum_i x_i^{p+1}} {\left(\sum_j x_j^p\right)\left(\sum_j x_j\right)}
\\
& = 1 - 4\, \frac{\sum_{1 \leq i < j \leq m} x_i x_j (x_i-x_j) \left(x_i^{p-1}-x_j^{p-1} \right)}
		{\left( \sum_j x_j \right)^2 \, \left(\sum_j x_j^p\right) }
\\
&\leq 1
\end{align*}
\end{IEEEproof}

\vspace{0.1in}
\begin{lemma}\label{lem:psd1}
Let ${\mathbf{a}}\in\mathbb{R}^m$ that satisfies $\sum_{i=1}^m a_i=1$ and $\sum_{i=1}^m a_i^2\leq 1$.
Then for any nonzero $\mathbf{z}\in \mathbb R^m$ we have
\bqn
f(\mathbf{z}):=\sum_{i=1}^m z_i^2 + \sum_{i=1}^m z_i \, \sum_{i=1}^m a_iz_i & > & 0
\eqn
provided $m \leq 8$.
\end{lemma}
\begin{IEEEproof}
Given any $M$ let $Z_M:=\{{z}\mid \sum_{i=1}^m z_i=M\}$.  It then suffices to show
that, for every $M\in \mathbb R$, $f(z)>0$ for $z\in Z_M$.
Given any $M$, consider
\bq
\min_{\mathbf{z}\in Z_M}f(\mathbf{z})  & = & \min_{\mathbf{z}\in Z_M} \ \sum_{i=1}^m z_i^2 + M \sum_{i=1}^m a_i z_i
\label{eq:120}
\eq
Its Lagrangian is
\bqn
L(\mathbf{z},\mu)=\sum_{i=1}^m z_i^2+M \sum_{i=1}^m a_i z_i + \mu \left(\sum_{i=1}^m z_i-M \right)
\eqn
where $\mu$ is the Lagrange multiplier.
Setting $\partial L/\partial z_i=0$ for all $1\leq i\leq m$ and substitute it
into $\sum_{i=1}^m z_i=M$, we obtain the unique minimizer given by
$\mu=-3M/m$ and $z_i=\frac{M}{2}(\frac{3}{m}-a_i)$.
Then
\bqn
\min_{\mathbf{z}\in Z_M}f(\mathbf{z}) & = & \frac{M^2}{4}\left(\frac{9}{m}-\sum_{i=1}^m a_i^2\right)
	\ \geq \ \frac{M^2}{4}\left(\frac{9}{m}-1\right)
\eqn
Hence, when $M\neq 0$, $\min_{\mathbf{z}\in Z_M}f(\mathbf{z})>0$ if $n<9$.
When $\mathbf{z}$ is nonzero but $M=0$, then $f(\mathbf{z})>0$ from \eqref{eq:120}.
\end{IEEEproof}

\section{Proof of Lemma \ref{lem:fluctuation}}\label{app:fluctuation}

By the definition of $D_k(A_k)$,  we have
\begin{align*}
\mathbb{E}\left[D_k(A_k)\mid\sum_{i,j}a_{ij}\geq1\right]=&d_k\mathbb{P}\left(\sum_{j}a_{kj}\geq 1\mid \sum_{i,j}a_{ij}\geq 1\right)\\
=&d_k\frac{\mathbb{P}(\sum_{j}a_{kj}\geq 1)}{\mathbb{P}(\sum_{i,j}a_{ij}\geq 1))}\\
=&d_k\frac{q_k|A_k|}{\sum_{i}q_i|A_i|}+o\left(\sum_{i}q_i\right),
\end{align*}
where the last equality follows from the independence of $a_{ij}$ and $\mathbb{P}(\sum_{j}a_{kj}\geq 1)=1-(1-q_k)^{|A_k|}=|A_k|q_k+o(q_k)$, $\mathbb{P}(\sum_{ij}a_{ij}\geq 1)=1-\prod_{i}(1-q_i)^{|A_i|}=\sum_{i}|A_i|q_i+\sum_{i}o\left(q_i\right)$. Thus,
\bqn
\mathbb{E}\left[\sum_{i}D_k(A_k)\mid\sum_{i,j}a_{ij}\geq1\right]=\frac{\sum_{k}d_kq_k|A_k|}{\sum_{k}q_k|A_k|}+o\left(\sum_{k}q_k\right).
\eqn

\begin{IEEEbiography}[{\includegraphics[width=1in,height=1.25in,clip,keepaspectratio]{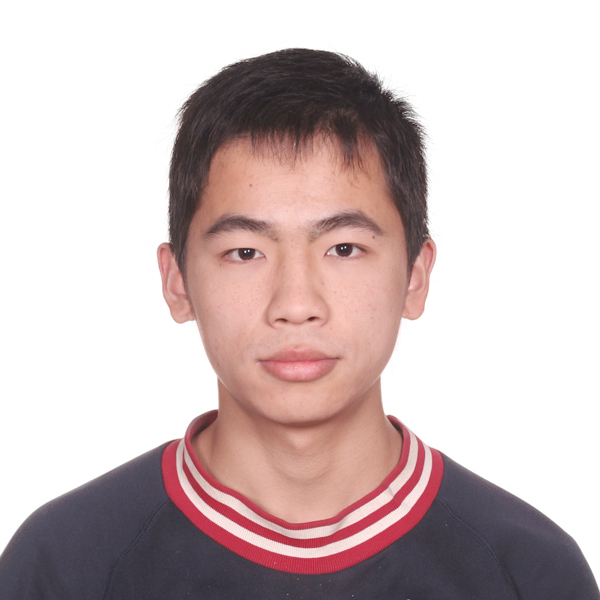}}]{Qiuyu Peng}
received his B.S. degree in Electrical Engineering from Shanghai Jiaotong University, China in 2011. He is currently working towards the Ph.D. degree in Electrical Engineering at California Institute of Technology, Pasadena, USA.

 His research interests are in the  distributed optimization and control for power system and communication networks.
\end{IEEEbiography}

\begin{IEEEbiography}[{\includegraphics[width=1in,height=1.25in,clip,keepaspectratio]{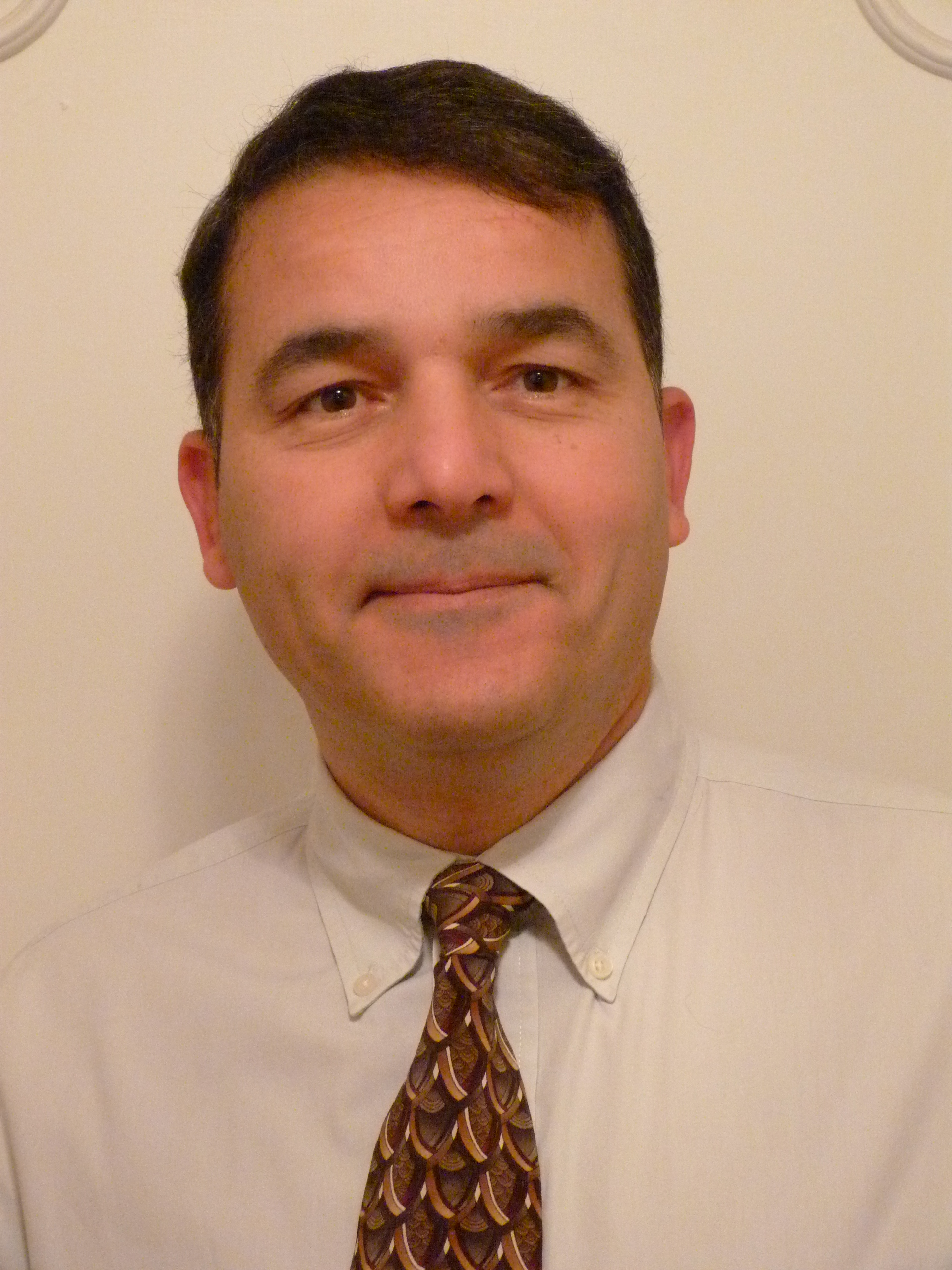}}]{Anwar Walid}
Anwar Walid is a Distinguished Member of Technical Staff with the Mathematics of Network Systems Research,  Bell Labs, Murray Hill, New Jersey.  He received a B.S. degree in electrical and computer engineering from Polytechnic of New York University, and a Ph.D. in electrical engineering from Columbia University, New York. He holds ten patents on computer and communication networks and systems. He received the Best Paper Award from ACM Sigmetrics, IFIP Performance. He has contributed to the Internet Engineering Task Force (IETF) with RFCs. He served as associate editor of the IEEE/ACM Transactions on Networking (ToN). He is associate editor of the IEEE/ACM Transactions on Cloud Computing and the IEEE Network Magazine. He was co-Chair of the Technical Program Committee of IEEE INFOCOM 2012. Dr. Walid is an IEEE Fellow and an elected member of Tau Beta Pi National Engineering Honor Society and the IFIP Working Group 7.3.
\end{IEEEbiography}

\begin{IEEEbiography}[{\includegraphics[width=1in,height=1.25in,clip,keepaspectratio]{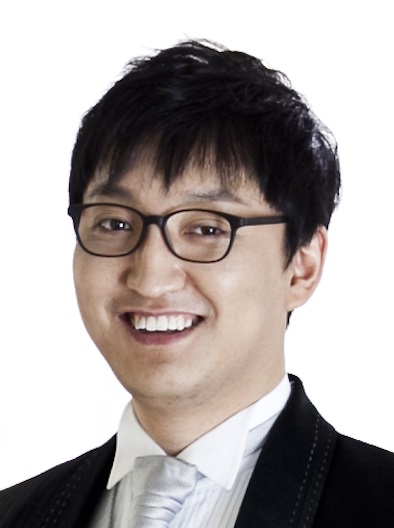}}]{Jaehyun Hwang}
Jaehyun Hwang (M'10) received the B.S. degree in computer science from The Catholic University of Korea, Korea in 2003, and the M.S. and Ph.D. in computer science from Korea University, Seoul, Korea in 2005 and 2010, respectively. Since September 2010, he has been with Bell Labs, Alcatel-Lucent, Seoul, Korea as a Member of Technical Staff. His current research interests include data center protocols, software-defined networking, multipath TCP, and HTTP adaptive streaming.
\end{IEEEbiography}

\begin{IEEEbiography}[{\includegraphics[width=1in,height=1.25in,clip,keepaspectratio]{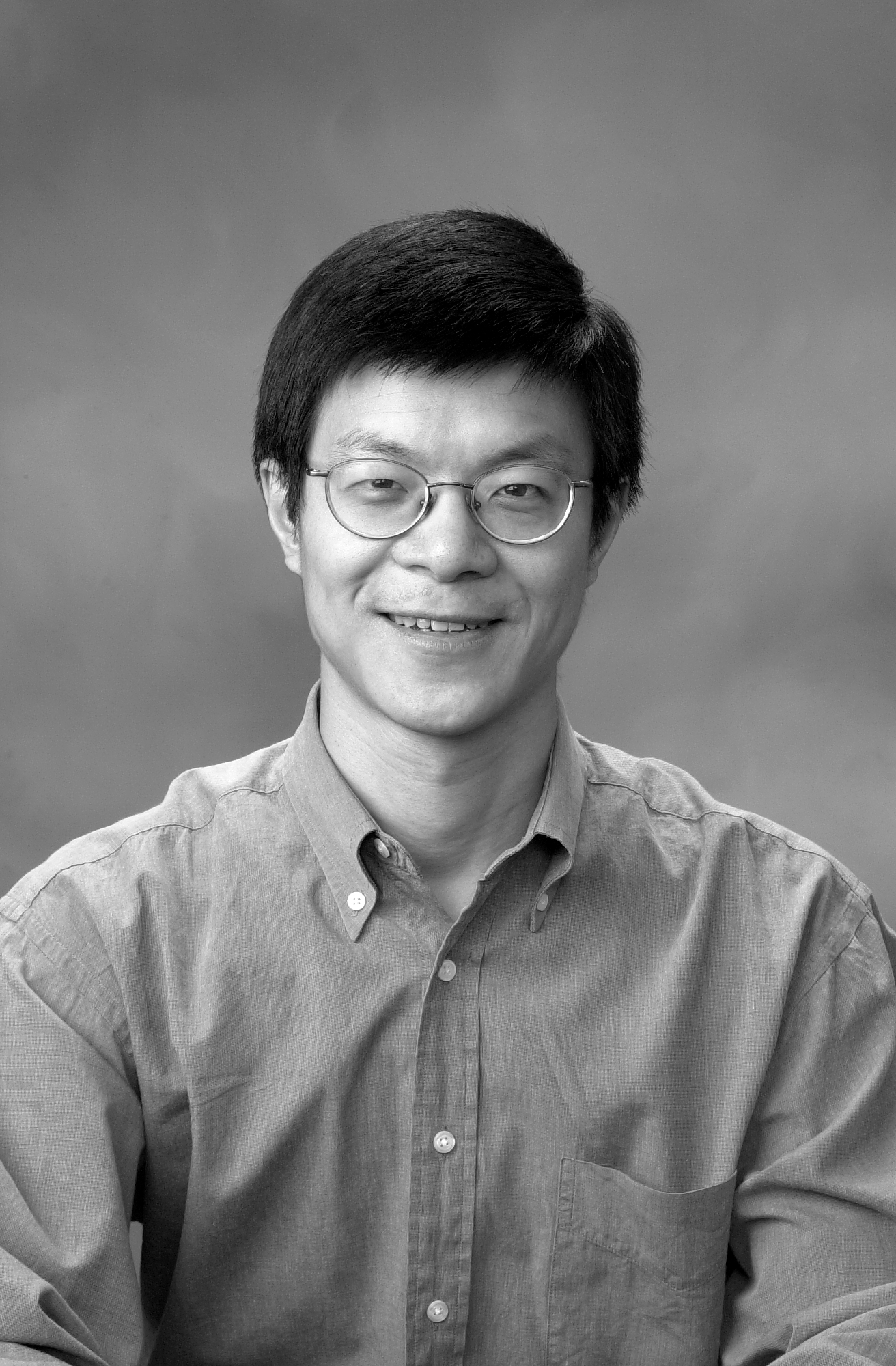}}]{Steven H. Low}
 (F'08)
is a Professor of the Department of Computing \& Mathematical
Sciences and the Department of Electrical Engineering at Caltech.
Before that, he was with AT\&T Bell Laboratories, Murray Hill, NJ, 
and the University of Melbourne, Australia.   He was a co-recipient of 
IEEE best paper awards, the R\&D 100 Award, and an Okawa Foundation 
Research Grant.  He is on the Technical Advisory Board of Southern 
California Edison and was a member of the Networking and Information 
Technology Technical Advisory Group for the US President's
Council of Advisors on Science and Technology (PCAST) in 2006.
He is a Senior Editor of the IEEE Transactions on Control of Network Systems
and the IEEE Transactions on Network Science \& Engineering,
is on the editorial boards of NOW Foundations and Trends in Networking, 
and in Electric Energy Systems, as well as Journal on Sustainable Energy,
Grids and Networks.
He received his B.S. from Cornell and PhD from Berkeley, both in EE.

\end{IEEEbiography}

\end{document}